\documentclass[twocolumn]{aastex63}
\received{July 1, 2020}
\revised{August 1, 2020}
\accepted{\today}
\submitjournal{ApJ}

\shortauthors{Tsuge et al.}

\usepackage{color}

\usepackage{multirow}

\begin{document}

\title{Active star formation across the whole Large Magellanic Cloud \\triggered by tidally-driven colliding H{\sc i} flows}

\author{Kisetsu Tsuge}
\affil{Department of Physics, Nagoya University, Furo-cho, Chikusa-ku, Nagoya 464-8601, Japan; tsuge@a.phys.nagoya-u.ac.jp}

\author{Hidetoshi Sano}
\affil{Department of Physics, Nagoya University, Furo-cho, Chikusa-ku, Nagoya 464-8601, Japan; tsuge@a.phys.nagoya-u.ac.jp}
\affil{Institute for Advanced Research, Nagoya University, Furo-cho, Chikusa-ku, Nagoya 464-8601, Japan}


\author{Kengo Tachihara}
\affil{Department of Physics, Nagoya University, Furo-cho, Chikusa-ku, Nagoya 464-8601, Japan; tsuge@a.phys.nagoya-u.ac.jp}

\author{Kenji Bekki}
\affiliation{ICRAR, M468, The University of Western Australia, 35 Stirling Highway, Crawley Western Australia 6009, Australia}

\author{Kazuki Tokuda}
\affiliation{Department of Physical Science, Graduate School of Science, Osaka Prefecture University, 1-1 Gakuen-cho, Naka-ku, Sakai, Osaka 599-8531, Japan}
\affiliation{National Astronomical Observatory of Japan, Mitaka, Tokyo 181-8588, Japan}

\author{Tsuyoshi Inoue}
\affil{Department of Physics, Nagoya University, Furo-cho, Chikusa-ku, Nagoya 464-8601, Japan; tsuge@a.phys.nagoya-u.ac.jp}

\author{Norikazu Mizuno}
\affiliation{National Astronomical Observatory of Japan, Mitaka, Tokyo 181-8588, Japan}

\author{Akiko Kawamura}
\affiliation{National Astronomical Observatory of Japan, Mitaka, Tokyo 181-8588, Japan}

\author{Toshikazu Onishi}
\affiliation{Department of Physical Science, Graduate School of Science, Osaka Prefecture University, 1-1 Gakuen-cho, Naka-ku, Sakai, Osaka 599-8531, Japan}

\author{Yasuo Fukui}
\affil{Department of Physics, Nagoya University, Furo-cho, Chikusa-ku, Nagoya 464-8601, Japan; tsuge@a.phys.nagoya-u.ac.jp}
\affil{Institute for Advanced Research, Nagoya University, Furo-cho, Chikusa-ku, Nagoya 464-8601, Japan}



\begin{abstract}


The galactic tidal interaction is a possible mechanism to trigger the active star formation in galaxies. Recent analyses using the H{\sc i} data in the Large Magellanic Cloud (LMC) proposed that the tidally driven colliding H{\sc i} flows, induced by the galactic interaction with the Small Magellanic Cloud (SMC), triggered high-mass star formation in the southeastern H{\sc i} ridge, including R136 and $\sim$400O/WR stars, and the galactic center region hosting the N44 region. This study performed a comprehensive H{\sc i} data analysis across the LMC and found that two H{\sc i} velocity components defined in the early studies (L- and D- components) are quasi-ubiquitous with signatures of interaction dynamically toward the other prominent H{\sc ii} regions, such as N11 and N79. We characterize the intimidate velocity range (I-component) between the two components as the decelerated gas by momentum conservation in the collisional interaction. The spatial distributions of the I-component and those of the O/WR stars have good agreements with each other whose fraction is more than $\sim$70\% at a scale of $\sim$15 pc, which is significantly smaller than the typical GMC size. 
Based on the results of our new simulations of the LMC-SMC interaction, we propose  that the interaction about 0.2 Gyr ago induced efficient infall of gas from the SMC to the LMC and consequently ended up with recent formation of high-mass stars due to collisions of H{\sc i} gas in the LMC.
The new numerical simulations of the gas dynamics successfully reproduce the current distribution of the L-component. This lends theoretical support for the present picture.
\end{abstract}


\section{Introduction} \label{sec:intro}

\subsection{Active star formation induced by the galactic tidal interaction}\label{sec:1.1}

Starburst is a key to understand the galaxy evolution and star formation history of the Universe. Many {early} studies suggest that tidal perturbations from nearby companions \citep{1988A&A...203..259N}, galactic interactions, mergers \citep[e.g., ][]{1988ApJ...325...74S, 2008MNRAS.388L..10B}, or cold gas accretion from the intergalactic medium \citep[e.g., ][]{1987ApJ...322L..59S} are possible mechanisms triggering the burst of star formation. \cite{1998Natur.395..859G} also show that the galaxy mergers have excess infrared luminosity compared with the isolated galaxies, lending support for the active star formation triggered by galaxy interactions. Recently, \cite{2014A&A...566A..71L}   investigated 18 starburst dwarf galaxies using H{\sc i} data. They found that starburst dwarf galaxies have more asymmetric H{\sc i} morphologies than typical dwarf irregulars, and most of starburst dwarf galaxies ($\sim$80 \%) are interacting galaxies which have at least one potential companion within 200 kpc. Thus, these previous works suggest that some external mechanism induced by galactic interaction triggers the starburst. In addition, numerical simulations of galaxy interactions show that interactions/mergers between gas-rich dwarfs formed irregular blue compact dwarfs (BCDs) including IZw 18, which hosts starburst \citep{2008MNRAS.388L..10B}. BCDs are usually low metallicity (0.1$\geq$$Z$/$Z_{\rm \odot}$$\geq$0.02) and is similar to the environment where the first stars formed in the early universe \citep{1972ApJ...173...25S}. Therefore, revealing the triggering mechanism of active star formation in dwarf galaxies has a potential to promote understanding of the origin of starburst in the early Universe. Most of the interacting galaxies are, however, distant, and it is difficult to resolve individual clouds and investigate their physical properties in detail{, except for recent ALMA studies toward nearby interacting system, such as the Antennae Galaxies (Tsuge et al. 2020a, b)}.

In the present study, we focus on the Large Magellanic Cloud (LMC). The LMC is one of the nearest interacting dwarf galaxies \citep[distance 50$\pm$1.3 kpc;][]{2013Natur.495...76P}, and almost face-on with an inclination of $\sim$20--30 deg. \citep[e.g., ][]{1972ApJ...173...25S}. {These condition enable us to} observe the whole galaxy uniformly without contamination. There are many active star-forming regions over the LMC, which have been intensively studied (N79: e.g., Ochsendorf et al. 2017; Nayak et al. 2019, N159: Chen et al. 2010; Saigo et al. 2017; Fukui et al. 2019; Tokuda et al. 2019, N11: e.g., Walborn \& Parker 1992; Celis Pena et al. 2019 , N44; e.g., Chu et al. 1993; Chen et al. 2009, N206: e.g.,Romita et al. 2010, N51: e.g., Lucke \& Hodge 1970; Chu et al. 2005, N105:e.g., Epchtein et al. 1984; Oliveira et al. 2006, N113: e.g., Brooks \& Whiteoak 1997, N120: e.g., Lucke \ Hodge 1970; , N144; e.g., Lortet \& Testor 1988 etc. catalogued by Henize 1956). The mean metallicity of the LMC is approximately half of the solar metallicity \citep[0.3--0.5 $Z_{\rm \odot}$; ][]{1992ApJ...384..508R, 1997macl.book.....W}, 
which is close to the mean metallicity of the interstellar medium during the time of peak star formation \citep[redshift z$\sim$1.5; ][]{1999ApJ...522..604P}. Thus, LMC is optimal laboratory {in} investigating the active star formation mechanism through a wide {spatial-dynamic} range from a galactic-scale (kpc scale) down to a cluster scale (10--15 pc). 

\subsection{Previous studies of the LMC}\label{sec:1.2}
\begin{figure}[]
\begin{center}
\includegraphics[width=\linewidth,clip]{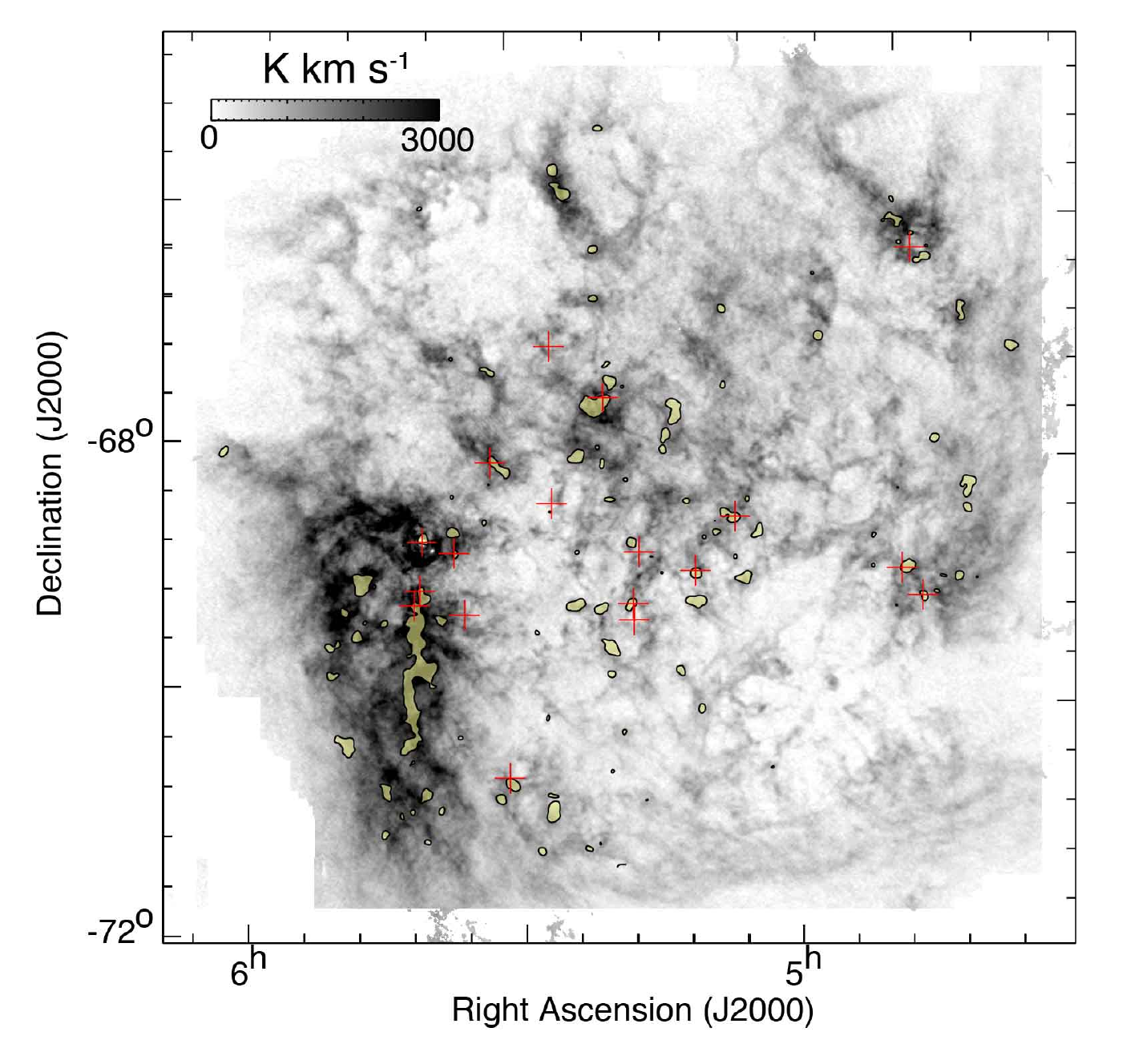}
\end{center}
\caption{The H{\sc i} integrated intensity maps with the integration velocity range of $V_{\rm offset}$=$-$101.1--29.9 km s$^{-1}$.  The yellow shaded areas are the regions where $^{12}$CO($J$=1--0) intensity obtained with NANTEN is greater than 1.5 $\sigma$ (1.48 K km s$^{-1}$) with the same velocity range. Plus signs show the positions of luminous H{\sc ii} regions (S$_{H\alpha}$ $>$ 1$\times$10$^{4}$ erg cm$^{-2}$ s$^{-1}$ sr$^{-1}$; Ambrocio-Cruz et al. 2016). }  
\label{fig1}
\end{figure}

In the context of the galactic tidal interaction, \cite{1990PASJ...42..505F} proposed that perturbation of the H{\sc i} gas induced by galactic tidal interaction between the LMC and the Small Magellanic Cloud (SMC) possibly triggered massive cluster formation. This model is supported by subsequent numerical simulations \citep{2007PASA...24...21B, 2014MNRAS.443..522Y}. According to these theoretical studies, the LMC and the SMC had a close encounter about 0.2 Gyr ago, and perturbed the gas and formed the current highly asymmetric H{\sc i}/CO gas distribution in the LMC, which is similar to the results of numerical simulation done by \cite{2007PASA...24...21B} as shown in Figure \ref{fig1}. Particularly, H{\sc i}/CO gas are concentrated in the southeast of the LMC. In addition, 
\cite{2007MNRAS.381L..16B} also suggested that metal poor gas have continued to flow from the SMC to the LMC at the velocity of 50--100 km s$^{-1}$ since 0.2 Gyr ago. This model indicates that the H{\sc i} gas consists of two velocity components whose velocity difference is $\sim$50 km s$^{-1}$; one is the H{\sc i} gas extending over the whole disk of the LMC named D-component and the other more locally distributed H{\sc i} gas named L-component \citep{1992A&A...263...41L} which has lower velocity relative to the D-component. \cite{1992A&A...263...41L} however did not mention the predicted of the above theoretical studies.

\begin{figure*}[htbp]
\begin{center}
\includegraphics[width=\linewidth]{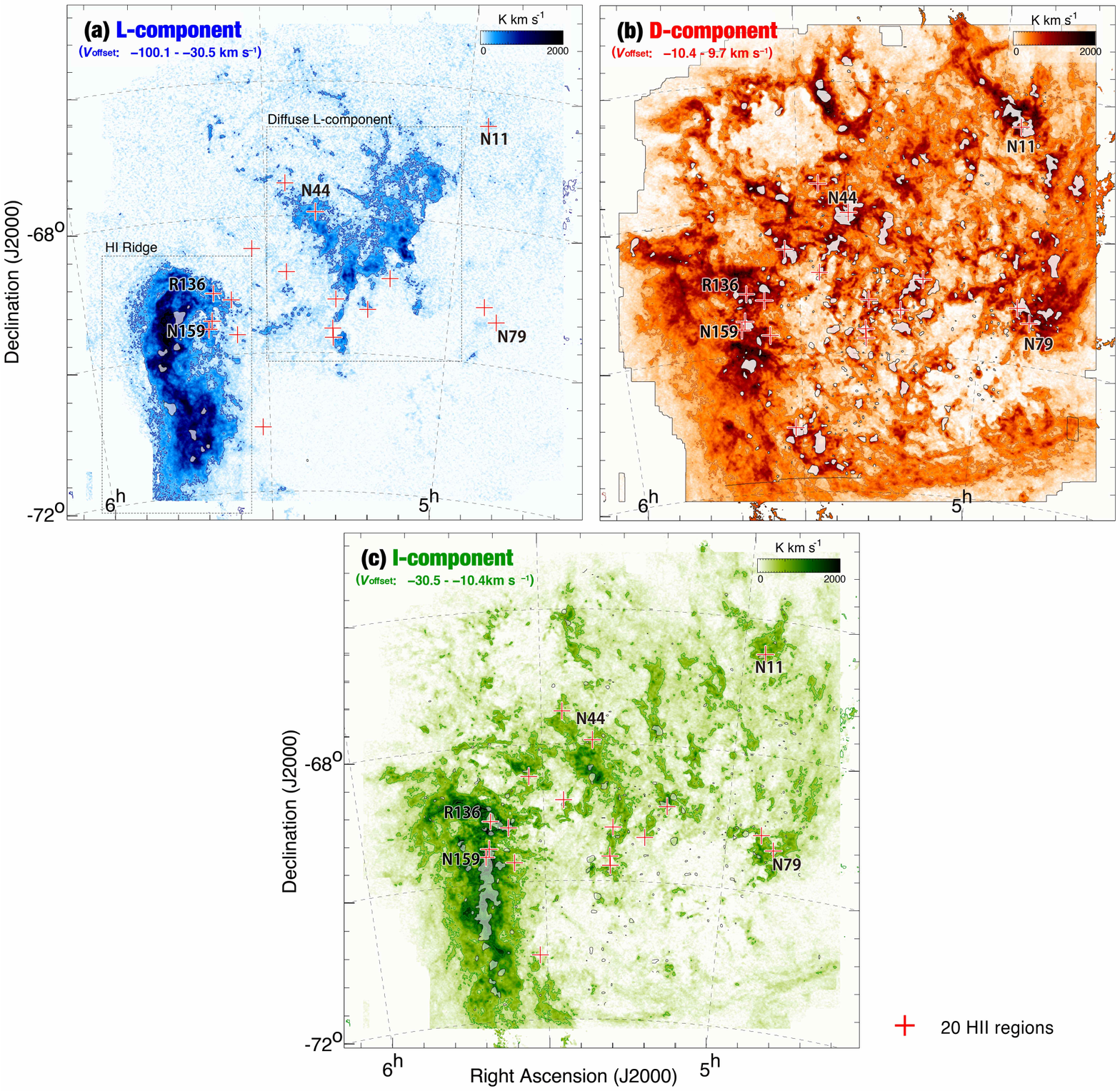}
\end{center}
\caption{The H{\sc i} integrated intensity maps of (a) the L-component, (b) the D-component, and (c) the I-component. The integration velocity range is  $V_{\rm offset}$=$-$101.1--$-$30.5 km s$^{-1}$ for the L-component, $V_{\rm offset}$=$-$10.4--9.7 km s$^{-1}$ for the D-component, and $V_{\rm offset}$=$-$30.5--$-$10.4 km s$^{-1}$ for the I-component. The lowest contour level and intervals are 192 K km s$^{-1}$ (15 $\sigma$) and 769 K km s$^{-1}$ (60 $\sigma$) for (a) and 279 K km s$^{-1}$ (40 $\sigma$) and 419 K km s$^{-1}$ (60 $\sigma$) for (b) and (c). The white shaded areas in three panels are the regions with $^{12}$CO($J$=1--0) intensity greater than 1.5 $\sigma$ with the same velocity range. $^{12}$CO($J$=1--0) Plus signs show the positions of luminous H{\sc ii} regions (S$_{H\alpha}$ $>$ 1$\times$10$^{4}$ erg cm$^{-2}$ s$^{-1}$ sr$^{-1}$; Ambrocio-Cruz et al. 2016). }  
\label{fig2}
\end{figure*}

Recently, it was found that the formation of LMC’s remarkable clusters R136 and N44 were induced by the galactic tidal interaction with the SMC (\citet{2017PASJ...69L...5F}; hereafter Paper I, \citet{2019ApJ...871...44T}; hereafter Paper II). The authors of Papers I and II analyzed the H{\sc i} data  \citep{2003ApJS..148..473K} of the whole LMC at a 1$\farcm$0 resolution corresponding to $\sim$15 pc at the distance of the LMC, and decomposed the L- and D-components (see Paper II for a detailed decomposition method). They found observational signatures as follows; i. the complementary spatial distribution between the L- and D- components and ii. The intermediate velocity components (hereafter I-component) connecting the two components in a velocity space, lending support for the collision of the L- and D-components as a formation mechanism of $\sim$400 high-mass stars including R136, N159, and N44. {Many numerical simulations elaborating cloud-cloud collision (e.g., Habe \& Ohta 1992; Anathpindika 2010; Takahira et al. 2014; 2018; Shima et al. 2018; Kobayashi et al. 2018; Sakre et al. 2020) have been performed, and synthetic observations by using the hydrodynamical numerical simulations of Takahira et al. (2014) showed that the i. complementary spatial distribution and ii. the intermediate velocity components connecting the two velocity components are observational signatures of cloud-cloud collision (Fukui et al. 2018).}
$V_{\rm offset}$ is defined as the relative velocity of the D-component ($V_{\rm offset}$ = $V_{\rm LSR}$$-$$V_{\rm D}$; $V_{\rm D}$ = the radial velocity of the D-component) in Paper I and II. The integration range is $V_{\rm offset}$: $-$100.1--$-$30.5 km s$^{-1}$ for the L-component; $V_{\rm offset}$: $-$30.5 -- $-$10.4 km s$^{-1}$ for the I-component; $V_{\rm offset}$: $-$10.4--9.7 km s$^{-1}$ for the D-component, and the method of determination of the integration ranges is shown in Paper II.  Figures \ref{fig2}a, \ref{fig2}b, and \ref{fig2}c show the spatial distribution of the L-, D-, and I-components, respectively. In addition, the dust abundance is found to decrease toward the colliding regions, which indicates that the H{\sc i} gas of the SMC is possibly mixed with the LMC gas. These results support the scenario that the collision of H{\sc i} gas was triggered by tidal interaction between the Magellanic Clouds.

{The previous studies have not subtracted the rotation velocity of the disk and the L- and D-components have not been decomposed over the whole LMC. We decomposed the L- and D-components over the whole LMC for the first time (Paper I and II). We defined the velocity ranges of these components by using position--velocity diagram of the northern part of the H{\sc i} Ridge region where the L- and D-components are clearly separated (see also Figure 2a).
We also made histogram of H{\sc i} gas toward the northern part of the H{\sc i} Ridge region shown in Figure \ref{fig16} of Appendix A. There are boundaries at $\sim$$-$30 km s$^{-1}$ and $\sim$$-$10 km s$^{-1}$ and it is consistent with the velocity ranges of the L- and D-components.}

\subsection{The I-component: a possibl{e} tracer of the colliding H{\sc i} flows}\label{sec:1.3}
\begin{figure}[htbp]
\begin{center}
\includegraphics[width=\linewidth]{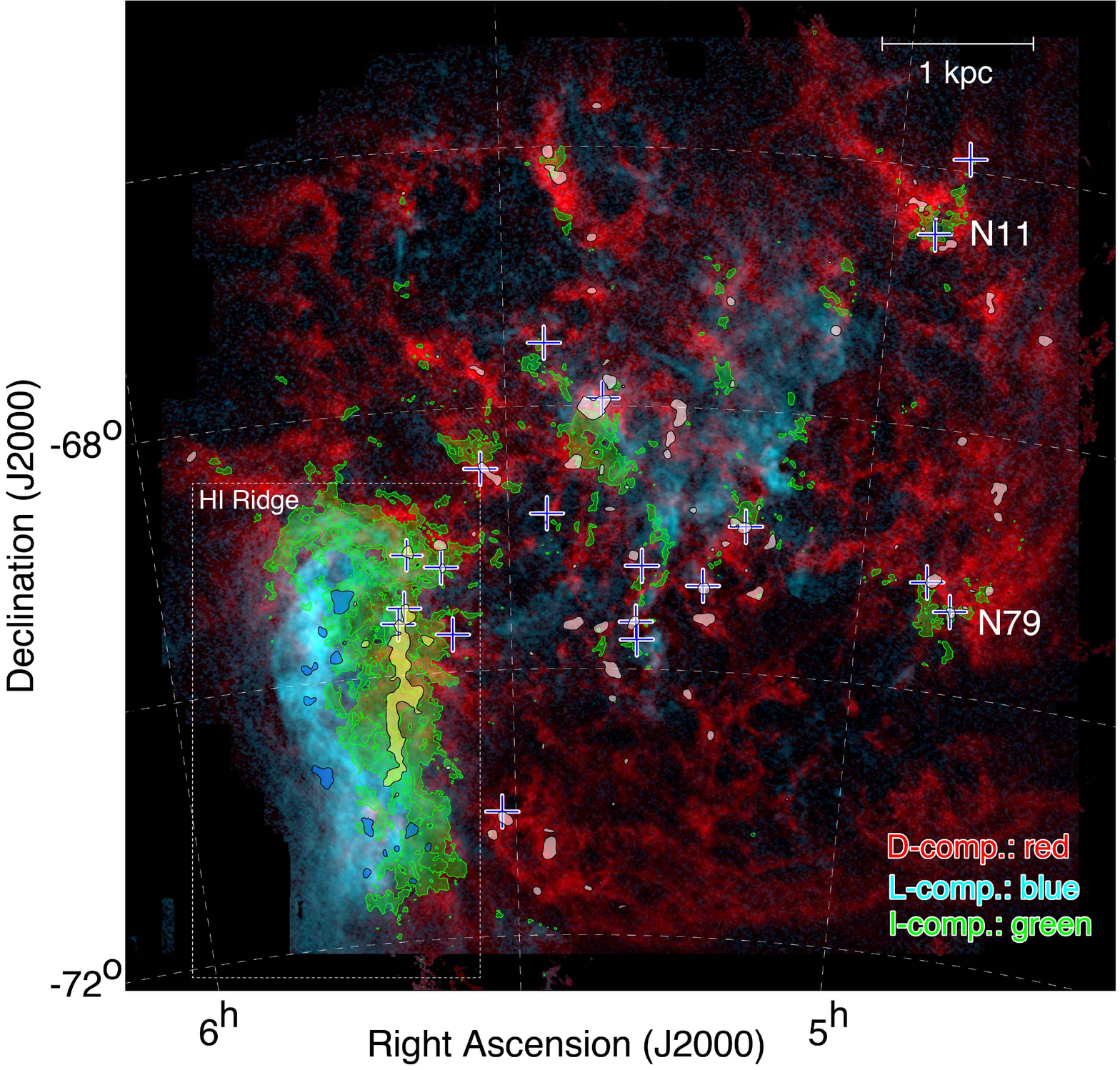}
\end{center}
\caption{H{\sc i} intensity map of the I-component by contours superposed on the L- and D-components image. The velocity ranges of three components are the same as in Figure 1. The contour levels are 500 and 1000 K km s$^{-1}$. Plus signs show the positions of luminous H{\sc ii} regions (S$_{H\alpha}$ $>$ 1$\times$10$^{4}$ erg cm$^{-2}$ s$^{-1}$ sr$^{-1}$; S$_{H\alpha}$ $>$ 1$\times$10$^{4}$ erg cm$^{-2}$ s$^{-1}$ sr$^{-1}$;. The shaded areas are the the regions where $^{12}$CO($J$=1--0) intensity obtained with NANTEN is greater than 1.5 $\sigma$ (1.48 K km s$^{-1}$) with the same velocity range. Those in yellow and blue indicate the Molecular Ridge and the CO-Arc, respectively.}  
\label{fig3}
\end{figure}

We defined the I-component whose velocity whose velocity is intermediate between the L- and D-components in Paper II. We interpreted that the I-component is decelerated gas in terms of momentum exchange in the collisional interaction of the L- and D-components. In fact, high-mass stars and molecular clouds were formed in the areas where the integrated intensity of the I-components is enhanced toward N44 and N159 (Paper II; \citet{2019ApJ...886...15T}, also see Figure \ref{fig7} of \citet{2019ApJ...886...14F}). The I-components is distributed over the whole LMC as shown in Figure \ref{fig2}c. Figure \ref{fig3} shows the spatial distributions of high integrated intensity areas (500 K km s$^{-1}$ $>$ $W$(H{\sc I}) ; assuming optically thin of the I-component overlaid with the L- and D-components. Spatial distribution of active H{\sc ii} regions \citep[][]{2016MNRAS.457.2048A} are similar to the strong I-component. This trend suggests that high-mass star formations are possibly triggering by the collision of H{\sc i} flows over the whole LMC in like manner of R136 and N44. Therefore, we aim to investigate relationship between the high-mass star formation and the I-component over the whole LMC and to reveal what percent of the high-mass stars were formed by collision of H {\sc i} flows. This paper is organized as follows. Section 2 summarizes the data sets, and Section 3 gives the results. The discussion is given in Section 4 discusses the formation mechanism of high-mass stars across the whole galaxy and Section 6 summarizes the paper.

\section{Data set} \label{sec:data}
\subsection{HI}\label{sec:2.1}
We used archival data of H{\sc i} 21 cm line emission of the whole LMC obtained with Australia Telescope Compact Array (ATCA) and Parkes telescope \citep{2003ApJS..148..473K}. The angular resolution of the combined H{\sc i} data is 60$\farcs$ (correspond to $\sim$15 pc at the distance of the LMC). The rms noise level is 2.4 K {at} a velocity resolution of 1.649 km s$^{-1}$. They combined the H{\sc i} data obtained by ATCA \citep{1998ApJ...503..674K} with those obtained with the Parkes multibeam receiver with a resolution of 14$\farcm$0--16$\farcm$0  \citep{1997PASA...14..111S}. The detailed descriptions of the observations are given by \citet{2003ApJS..148..473K}.

\subsection{CO}\label{sec:2.2}
$^{12}$CO($J$=1--0) data {obtained with the} NANTEN 4 m telescope \citep{1999PASJ...51..745F, 2008ApJS..178...56F, 2001PASJ...53..971M} are used for a large-scale analysis. This observation uniformly covered 6$^{\circ}$$\times$6$^{\circ}$ area including the whole optical extent of the LMC, and is suitable for the comparison with kpc scale H{\sc i} dynamics.  The half-power beam width is 2$\farcm$6 with grid spacing of 2$\farcm$0, and the velocity resolution is 0.65 km s$^{-1}$.

We also used $^{12}$CO($J$=1--0) data of the Magellanic Mopra Assessment \citep[MAGMA; ][]{2011ApJS..197...16W} for a small-scale analysis in each star-forming region. The angular resolution is 45” (correspond to 11 pc at the distance of the LMC), and the velocity resolution is 0.526 km s$^{-1}$. However, the MAGMA survey does not cover the whole LMC, and the observed area is limited toward the individual CO clouds detected by {the} NANTEN survey.

\subsection{H$\alpha$}\label{sec:2.3}
We used the H$\alpha$ data obtained by the Magellanic Cloud Emission-Line Survey \citep[MCELS; ][]{1999IAUS..190...28S}. The dataset was obtained with a 2048$\times$2048 CCD camera on the Curtis Schmidt Telescope at Cerro Tololo Inter-American Observatory. The angular resolution is $\sim$3$\farcs$--4$\farcs$ (correspond to $\sim$0.75--1.0 pc at a distance of the LMC). We also use the archival data of H$\alpha$ provided by the Southern H-Alpha Sky Survey Atlas \citep[SHASSA; ][]{2001PASP..113.1326G} in order to define the region where UV radiation is locally enhanced by star formation.

\section{Results}\label{sec:results}

\subsection{Spatial- and velocity-structures of HI gas at kpc scale}\label{sec:3.1}
We compare the distribution of I-component with major 20 star-forming regions over the whole LMC. Figures \ref{fig2}(a), \ref{fig2}(b), and \ref{fig2}(c) present distributions of the L-, D-, and I-components, and Figure \ref{fig3} is an overlay {map} of the three components. The L-component is composed of two extended emission. One is the H{\sc i} Ridge region located in the south east region. The H{\sc i} Ridge region includes two major elongated CO clouds, the Molecular Ridge and CO-Arc. The other is the diffuse component extending {toward} the northwest as defined in Paper II (hereafter diffuse L-component). The D-component is distributed over the whole LMC. The I-component is distributed along the western rim of the L-component of H{\sc i} Ridge and the southern rim of the diffuse L-component as shown in Figure \ref{fig3}. The I-component is also located toward the southern end of western tidal arm including N79 and the northern end of the western tidal arm including N11. Most of the major star-forming regions \citep{2016MNRAS.457.2048A} are located in the I-component. Molecular clouds also exhibit good spatial correlation between the I-component except for the CO-arc region.
Physical parameters of the L-, D-, and I-components are summarized in Table \ref{tab:mass}. We calculate mass of the H{\sc i} gas in the assumption that H{\sc i} emission is optically thin as follows, $N_{\rm H{\sc I}} = 1.8224\times10^{18}\int \Delta T_{\rm b}  \it dv \ \rm{[cm^{-2}] }$, where $T_{\rm b}$ is the observed H{\sc i} brightness temperature (K). The masses of atomic hydrogen and molecular hydrogen are 0.3$\times$10$^8$ $M_{\rm \odot}$ and 0.3$\times$10$^7$ $M_{\rm \odot}$ for the L-component, 1.8$\times$10$^8$ $M_{\rm \odot}$ and 2.0$\times$10$^7$ $M_{\rm \odot}$ for the D-component, and 0.8$\times$10$^7$ $M_{\rm \odot}$ and 0.9$\times$10$^7$ $M_{\rm \odot}$ for the I-component. We also derived the masses of the molecular clouds using the $W_{\rm CO}$--$N$(H$_{2}$) conversion factor \citep[$X_{\rm CO}$ = 7.0$\times$10$^{20}$ cm$^{-2}$ (K km s$^{-1}$)$^{-1}$; ][]{2008ApJS..178...56F}. We use the equation as follows: $N$(H$_{2}$)=$X_{\rm CO}$$\times$$W$($^{12}$CO($J$=1--0)), where $W$ is the integrated intensity of $^{12}$CO($J$=1--0) and $N$(H$_{2}$) is the column density of molecular hydrogen. 

\begin{deluxetable}{cccc}
\tablewidth{7.0cm}
\tablecaption{Mass of hydrogen gas}
\label{tab:mass}
\tablehead{\multicolumn{1}{c}{H{\sc i} } &M(H{\sc i})&$M$(H$_{2}$)  \\
\multicolumn{1}{c}{components}&{[}10$^{8}$$M_{\odot}${]}&{[}10$^{7}$$M_{\odot}${]}\\
\multicolumn{1}{c}{(1)} &(2)&(3)}
\startdata
L-component & 0.3& 0.3  \\
D-component & 1.8& 2.0 \\ 
I-component & 0.8& 0.9\\ 
\enddata
\end{deluxetable}

\begin{figure*}[htbp]
\begin{center}
\includegraphics[width=16cm,clip]{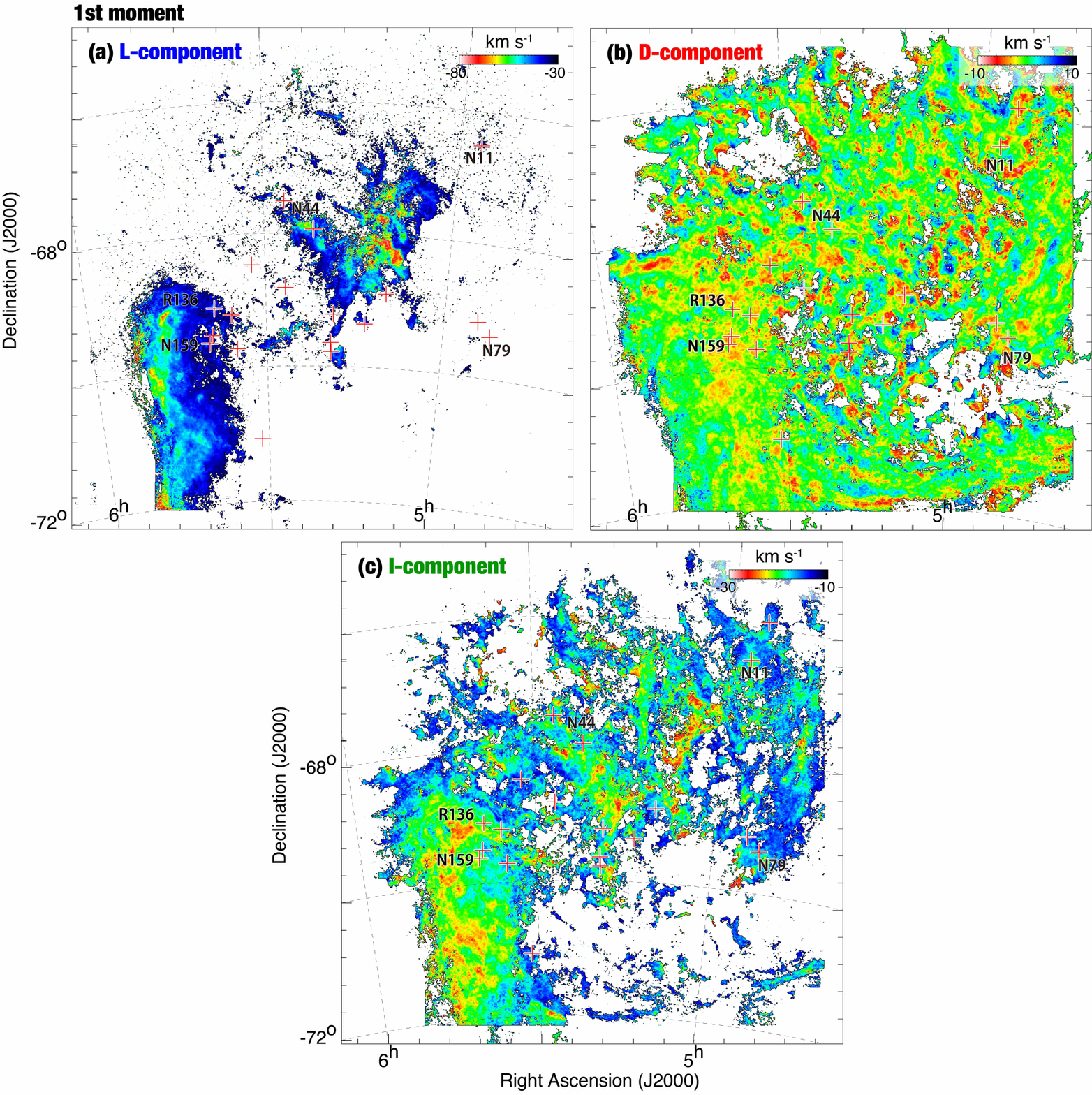}
\end{center}
\vspace*{-3cm}
\caption{1st moment map of H{\sc i} over the whole LMC. 1st moment is the intensity-weighted velocity following the equation of
$\Sigma$($I$ $\times$ $v$)/$\Sigma$($I$), where $I$ is intensity of emission and $v$ is velocity. Velocity ranges using for calculating momentum are $V_{\rm offset}$=$-101.1$--$-$30.5 km s$^{-1}$ (the L-component) for (a); $V_{\rm offset}$=$-10.4$--$-$9.7 km s$^{-1}$ (the D-component) for (b); $V_{\rm offset}$=$-30.5$--$-$10.4 km s$^{-1}$ (the I-component) for (c). The symbols are the same as in Figure 1.}  
\label{fig4}
\end{figure*}

Figure \ref{fig4} shows the 1st moment maps of the three components. 1st moment is the intensity-weighted velocity following the equation of $\Sigma$($I$ $\times$ $v$)/$\Sigma$($I$), where $I$ is intensity of emission and $v$ is $V_{\rm offset}$. Figure \ref{fig4}(a) shows the 1st moment of the L-component. For the H{\sc i} Ridge region, there is a velocity gradient from the east to the west. Typical velocity of the eastern side is $-$60 km s$^{-1}$, and the velocity is increasing to $-$30 km s$^{-1}$ in the western side. For the Diffuse L-component, there is a velocity gradient from the center to the edge of Diffuse L-component. Typical velocity of the center is $-$70 km s$^{-1}$ as shown by red, and the velocity is increasing to $-$30 km s$^{-1}$ in the edge as shown by blue. Figure \ref{fig4}(b) shows the 1st moment of the D-component. There is no drastic change of velocity. Figure \ref{fig4}(c) shows the 1st moment of the I-component. Velocity of the I-component is affected by the L-component. An average value is $\sim$$-$15 km s$^{-1}$, while toward the eastern side of the H{\sc i} Ridge and the edge of the Diffuse L-component, velocity decreases to about $-$25 km s$^{-1}$ as shown by red.

\subsection{Comparison of the I-component and high-mass stars}\label{sec:3.2}
\begin{figure*}[htbp]
\begin{center}
\includegraphics[width=\linewidth]{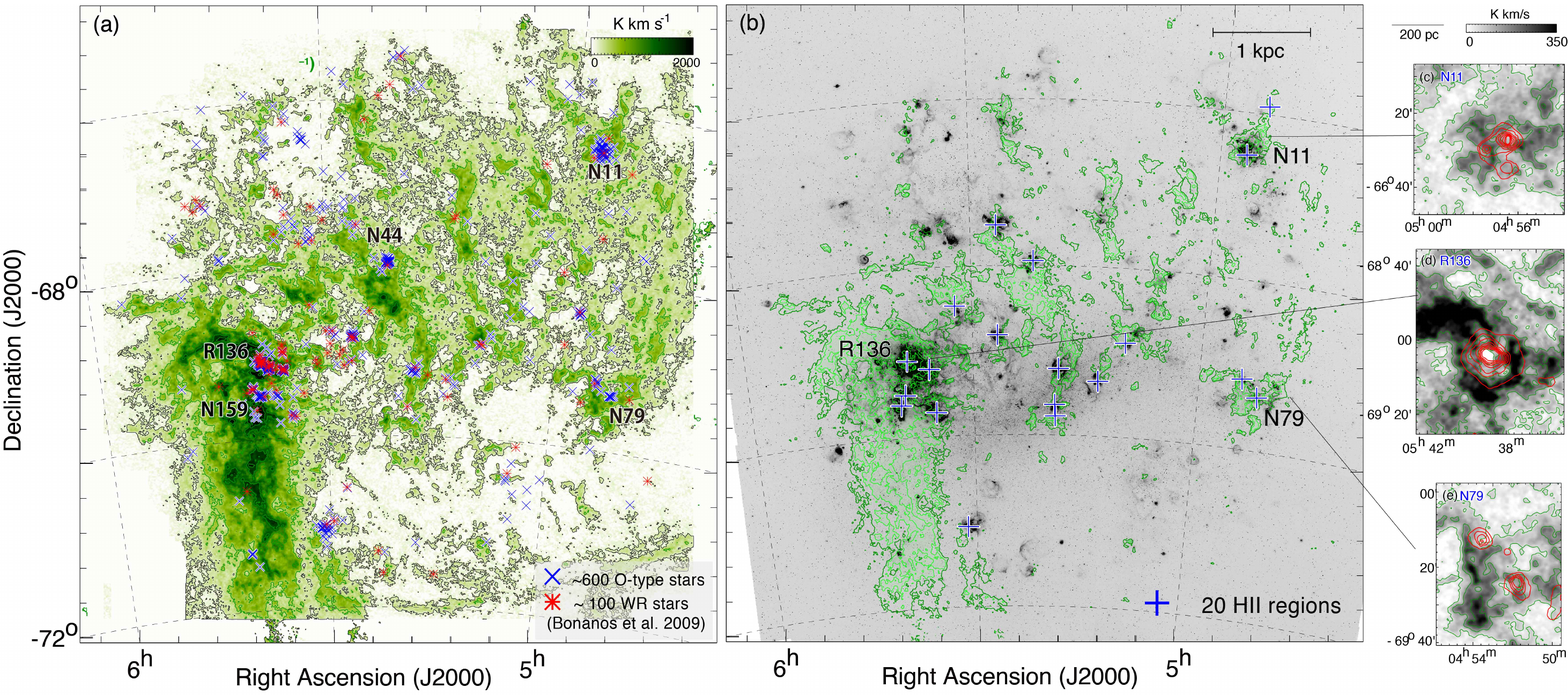}
\end{center}
\vspace*{-2cm}
\caption{(a) The H{\sc i} integrated intensity map of the I-component. The red asterisks and blue crosses indicate WR stars and O-type stars (Bonanos et al. 2009), respectively. (b) The H{\sc i} integrated intensity map of the I-component by contours superposed on H$\alpha$ image. The symbols are the same as in Figure 1. (c) (d) (e) show enlarged view toward N11, R136, and N79, respectively. Images and green contours show H{\sc i} distributions. Red contours show H$\alpha$ image. The integration range is $V_{\rm offset}$=$-15.6$--$-$12.3 km s$^{-1}$ for (a); $V_{\rm offset}$=$-18.8$--$-$15.6 km s$^{-1}$ for (b); $V_{\rm offset}$=$-15.6$--$-$12.3 km s$^{-1}$ for (c). The contour levels of H{\sc i} are 50, 150, and 250 K km s$^{-1}$. The contour levels of H$\alpha$ are 150, 300, 450, and 600 R. Detailed velocity channel maps are shown in Appendix B. }  
\label{fig5}
\end{figure*}

In order to investigate the physical relationship between the I-component and high-mass star formation, we compared in Figure \ref{fig5}(a) the special distributions of the I-component and $\sim$700 O-type/WR stars \citep{2009AJ....138.1003B}. Most of the O-type/WR stars are located toward the I-component. On the other hand, there are many regions where the I-component without high-mass stars such as the southern part of the H{\sc i} Ridge region or the southern part of the N44 region as shown in Figure \ref{fig5}(a). We will discuss  interpretation of this trend in section 4.3.

Figure \ref{fig5}(b) shows the distributions of the H$\alpha$ emission and the I-component. As shown in Figure \ref{fig3}, spatial correlation between the major H{\sc ii} regions and the I-component is seen at kpc scale. In addition, we compared detailed spatial distributions of H$\alpha$ emission and H{\sc i} gas toward R136, N11, and N79 at 10--100 pc scale. Figures \ref{fig5}(c), \ref{fig5}(d), and \ref{fig5}(e) show enlarged views of H{\sc i} toward N11, R136, and N79, respectively. These figures exhibit intensity depression toward H{\sc ii} region in the velocity range of the I-component. Detailed velocity channel maps toward N11, R136, and N79 are shown in Figures \ref{fig16}, \ref{fig17}, and \ref{fig18} of Appendix B.

\begin{figure}[htbp]
\begin{center}
\includegraphics[width=\linewidth]{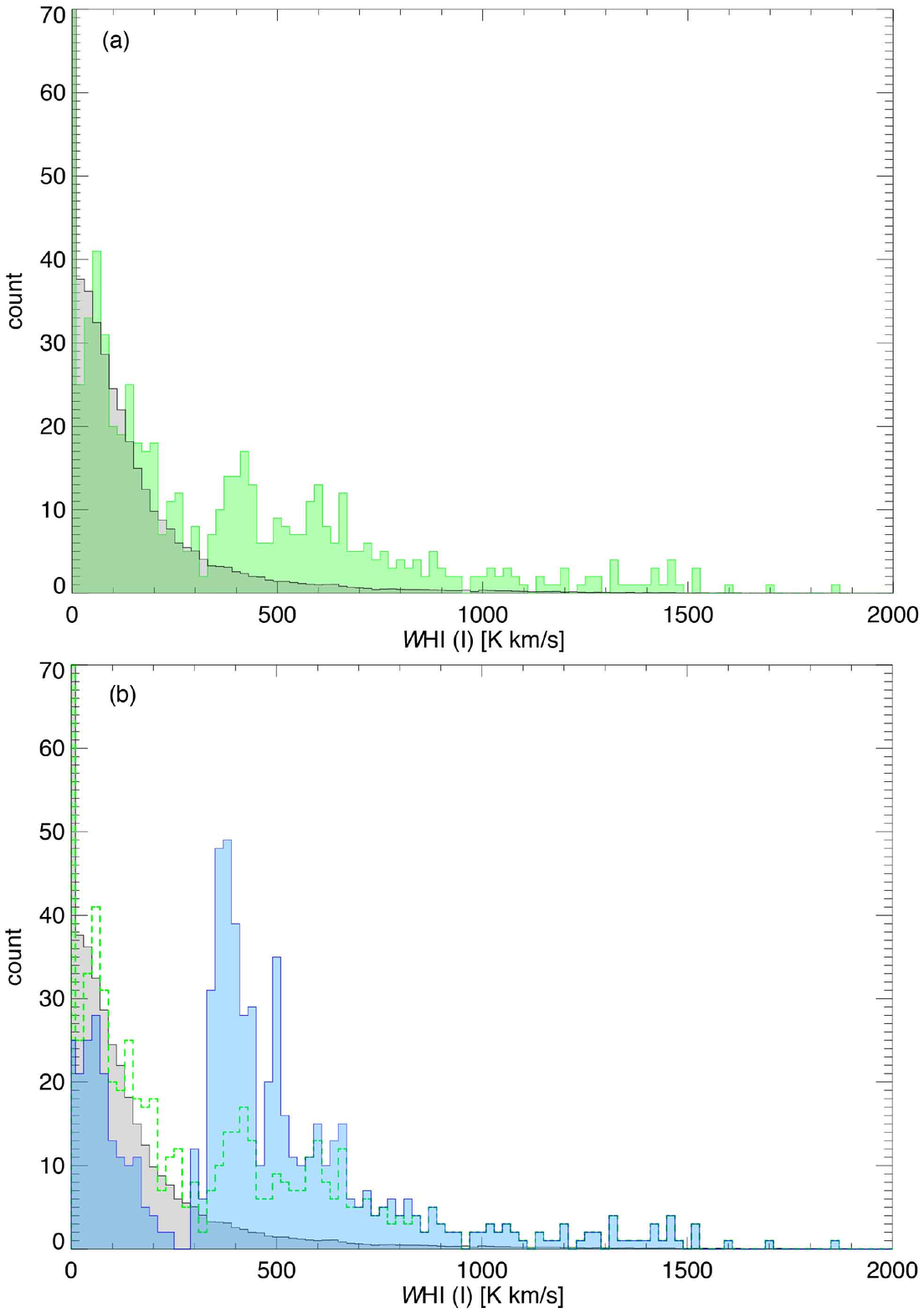}
\end{center}
\vspace*{-1cm}
\caption{(a) Histograms of integrated intensity of the I-component ($W$H{\sc i} (I)) at the positions of 697 O-type/WR stars cataloged by Bonanos et al. (2009) shown in green. Grey histogram shows the result expected if the same number of high-mass stars are distributed at random over the whole LMC. (b) Green and grey histograms are the same as in (a). Blue histogram shows a result considered the effect of Ha emission. Detailed method is shown in section section 3.2. }  
\label{fig6}
\vspace*{-0.5cm}
\end{figure}

In order to confirm the spatial correlation between the high-mass stars and the I-component, we produced a histogram of integrated intensity of the I-component (hereafter, $W_{\rm H{\sc i}}$ (I)) at the positions of 697 O-type/WR stars cataloged by \cite{2009AJ....138.1003B} as shown in Figure \ref{fig6}. It was found that $\sim$50\% of the O-type/WR stars are located at the positions where $W_{\rm H{\sc i}}$ (I) $>$ 300 K km s$^{-1}$ (green histogram of Figure \ref{fig6}(a)). This correlation cannot be due to chance coincidence, because the characteristics of the histogram is significantly different from what is expected for the case of a purely random distribution as shown by the grey histogram of Figure \ref{fig6}(a). However, the effect of the H$\alpha$ emission should be considered as mentioned in the previous paragraph. The H{\sc i} gas will be ionized by the stellar UV radiation, causing intensity decrease of the I-component. We therefore modified the values of $W_{\rm H{\sc i}}$ (I) when the following conditions i) and ii) are satisfied; i) $W_{\rm H{\sc i}}$ (I) is less than 300 K km s$^{-1}$ and ii) H$\alpha$ emission is higher than 500 dR are to be satisfied. We assumed $W_{\rm H{\sc i}}$ (I) from the H{\sc i} data around the H$\alpha$ emissions. If $W_{\rm H{\sc i}}$ (I) within 50 pc radius from high-mass star is higher than 300 K km s$^{-1}$, we adopted the value as new $W_{\rm H{\sc i}}$ (I) toward the star. 50 pc is the expected size of an inozied cavity around a star in 10 Myr when the velocity of ionization is roughly assumed to be 5 km s$^{-1}$ \citep{2018ApJ...859..166F}. Figure \ref{fig6}(b) shows the histogram corrected for the H{\sc i} decrease toward the H$\alpha$ emission. It was found that $\sim$70\% of O-type/WR stars are located at the positions where $W_{\rm H{\sc i}}$ (I) $>$ 300 K km s$^{-1}$ as shown by the blue histogram of Figure \ref{fig6}(b)) and the difference from the random case was more significant than that to Figure \ref{fig6}(a). {As shown in channel maps of the P--V diagrams of Appendix D, there is the I-component (bridge features) toward $\sim$70 \% of high-mass stars.}
Spatial distributions of O-type/WR stars correlated/uncorrelated with I-component are shown in Figure \ref{fig19} of Appendix C. {Most of 30\% uncorrelated high-mass stars are located inside of the SGSs and there are few bridge features (the I-component).}

\subsection{Spatial- and velocity-structures of HI gas toward individual region{s}}\label{sec:3.3}

As mentioned in section 3.2, it is suggested that most of the high-mass star formation is possibly related with the I-component. In this section, we investigate detailed spatial and velocity structures of H{\sc i} and CO toward the northern part of H{\sc i} Ridge, N11, and N79 as following the method of Paper I /II in order to investigate the possibility of high-mass star formation triggered by colliding H{\sc i} flows. N11 and N79 are located at the northern end and the southern end of the western tidal arm, respectively. 

\subsubsection{HI Ridge region}\label{sec:3.3.1}
\begin{figure*}[htbp]
\begin{center}
\includegraphics[width=\linewidth]{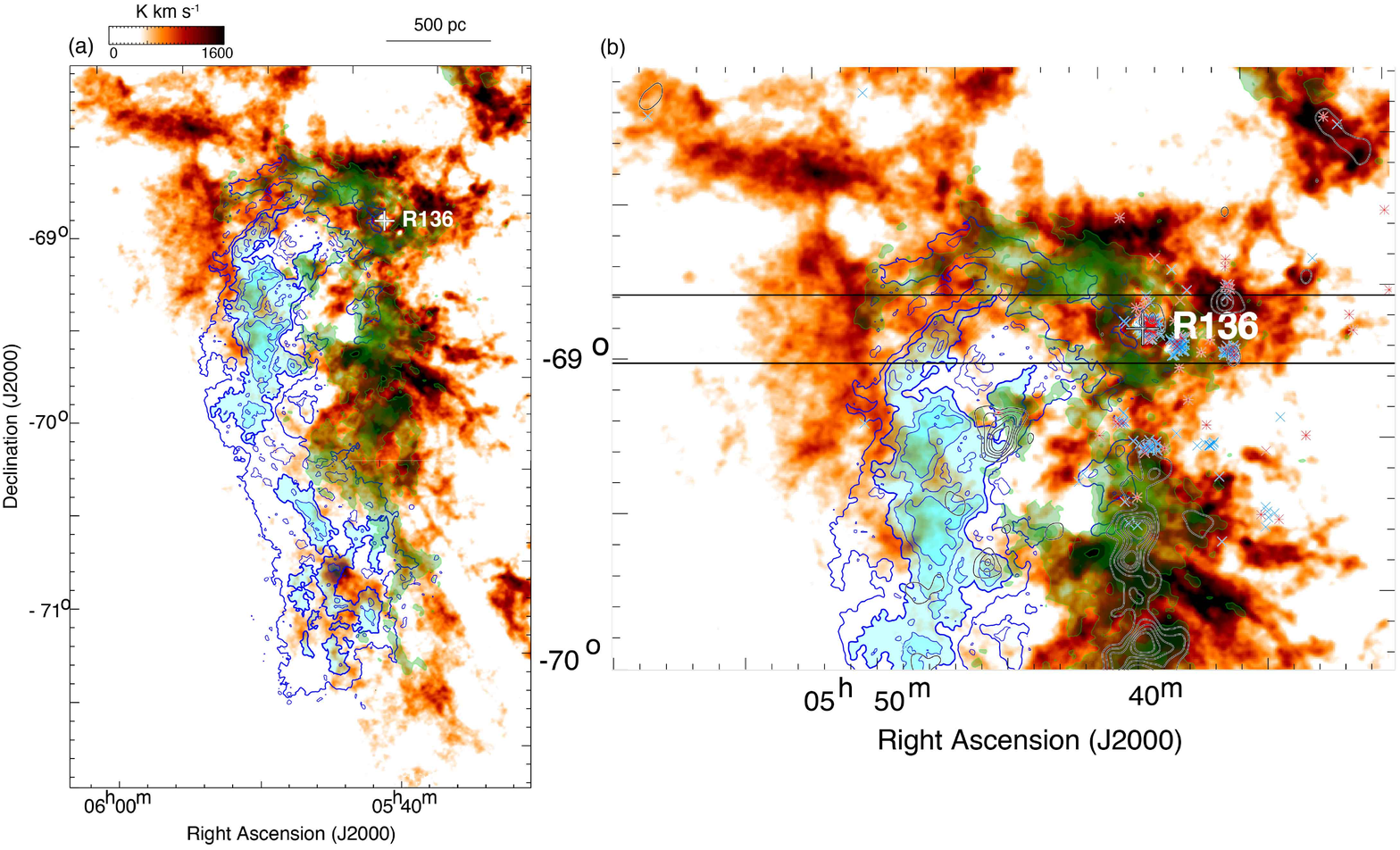}
\end{center}
\vspace*{-1cm}
\caption{(a) Intensity map of H{\sc i} consisting of three velocity components (the L-, I-, and D-components) of the H{\sc i} Ridge. The intensity map of the L-component by blue contours superposed on the D-component (red image). The contour levels are 500, 800, 1000, 1200, 1400, 1600, 1800, and 1900 K km s$^{-1}$. The green shaded area and contours of (a) and (b) indicate the I-component where integrated intensity is larger than 800 K km s$^{-1}$. The contour levels are 800, 1200, 1600, and 2000 K km s$^{-1}$. (b) Enlarged view of the northern part the H{\sc i} Ridge. The red asterisks and blue crosses indicate WR stars and O-type stars (Bonanos et al. 2009), respectively. The black contours indicate the distributions of $^{12}$CO($J$=1--0) obtained {with the} NANTEN telescope. The integration velocity range is $V_{\rm offset}$=$-101.1$--9.7 km s$^{-1}$. The lowest contour level and intervals are 1.48 K km s$^{-1}$ (1.5 $\sigma$).}  
\label{fig7}
\end{figure*}

We focus on the H{\sc i} Ridge region. In Paper I, we only discussed on high-mass star formation around R136, and did not investigate the physical properties of the I-component. Figure \ref{fig7}(a) shows spatial distributions of the L-, I-, and D-components toward the H{\sc i} Ridge region. The L-component has complementary distribution with the D-component. The D-component shows intensity depression toward the dense part of the L-component. The I-component is distributed from the north to the south of the H{\sc i} Ridge, and is spatially connected with the L- and D-component at Declinations of $\sim$$-$68$^d$45$^m$, $\sim$$-$69$^d$20$^m$, $\sim$$-$69$^d$45$^m$, and from $\sim$$-$70$^d$20$^m$ to $-$71$^d$00$^m$.

\begin{figure*}[htbp]
\begin{center}
\includegraphics[width=14cm]{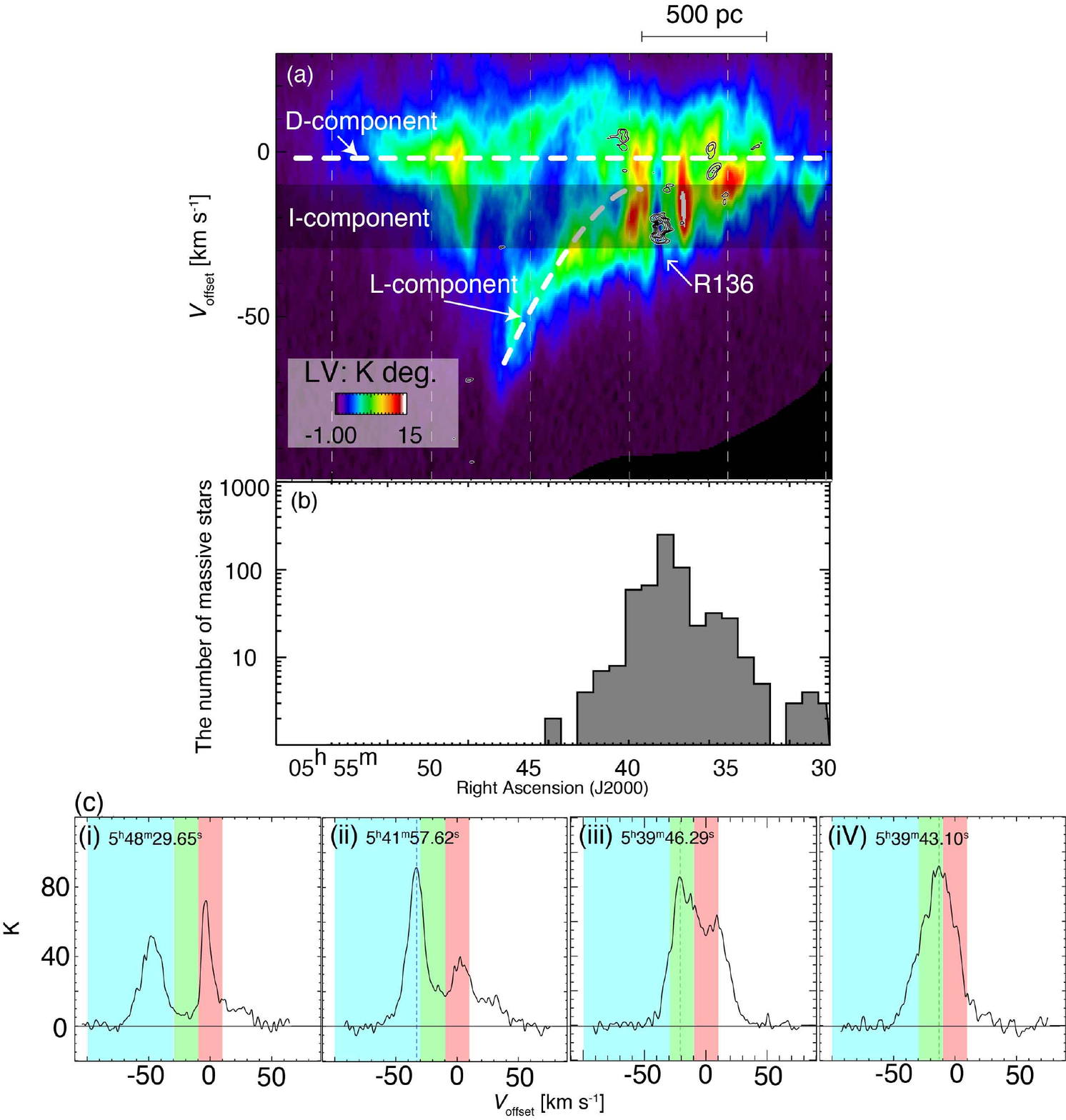}
\end{center}
\vspace*{-2cm}
\caption{(a) Right Ascension--velocity diagram toward the northern part of H{\sc i} Ridge. Integration range in Declination is from -69.01 deg. to -68.79deg. at R.A. of 6h 00m 00s as shown in black lines in Figure 7(b). The black contours indicate $^{12}$CO ($J$=1--0). The contour levels are 0.0133, 0.0171, 0.0209, 0.0247, and 0.0285 K deg. (b) Histogram of the high-mass stars toward the northern part of H{\sc i} Ridge as shown in Figure 7(b). The horizontal and vertical axes are R.A. and the number of massive stars located within each R.A. bin, respectively. (c) Typical spectra of H{\sc i} at the position of (i) (R.A., Dec.)=(5$^{h}$48$^{m}$29.65$^{s}$,-69$^{d}$23'3.49"), (ii) (R.A., Dec.)=(5$^{h}$41$^{m}$57.62$^{s}$,-69$^{d}$09'26.54"), (iii) (R.A., Dec.)=(5$^{h}$39$^{m}$46.29$^{s}$,-69$^{d}$09'42.95"), and (iV) (R.A., Dec.)=(5$^{h}$39$^{m}$43.10$^{s}$,-69$^{d}$00'20.42"). Blue, green, and red shaded ranges are $V_{\rm offset}$=$-101.1$--$-$30.5 km s$^{-1}$ (the L-component), $V_{\rm offset}$=$-30.5$--$-$10.4 km s$^{-1}$ (the I-component), and $V_{\rm offset}$=$-10.4$--$-$9.7 km s$^{-1}$ (the D-component), respectively.}  
\label{fig8}
\end{figure*}

As shown in Figure \ref{fig7}(b), the I-component exhibits {a} good spatial correlation with the molecular clouds (black contours) and high-mass stars (blue crosses and red asterisks) toward the northern part of the H{\sc i} Ridge region. In addition, velocity structure of northern part of the H{\sc i} Ridge is shown in Figure \ref{fig8}(a). The integration range in Dec. is from $-$69.01 deg. to $-$68.79 deg. at R.A. = 6$^h$00$^m$00$^s$. In the western side (R.A. $>$ 5$^h$45$^m$), the L- and D-components are clearly separated from each other with velocity difference of $\sim$50 km s$^{-1}$. The velocity difference of the L- and D-components becomes smaller as the R.A. becomes more from the west to the east as shown in the white dashed line of Figure \ref{fig8}(a). This trend is also seen in the typical spectrum of H{\sc i} as shown in Figure \ref{fig8}(c). Moreover, the intensity of the H{\sc i} gas is enhanced by merging of the L- and D-component, and molecular clouds are formed at the R.A. from 5$^h$35$^m$ to 5$^h$40$^m$ where young massive cluster R136 is located. We interpret that by merging of the L- and D-components the molecular clouds are formed. At the R.A. of R136 (5$^h$38$^m$42.396$^s$), H{\sc i} gas is ionized by high-mass stars in a velocity range of the I-component. Figure \ref{fig8}(b) shows a histogram of the number of high-mass stars. Most of the high-mass stars are also formed at the same R.A. from 5$^h$35$^m$ to 5$^h$40$^m$.

\subsubsection{LHA 120-N11}\label{sec:3.3.2}
\begin{figure*}[htbp]
\begin{center}
\includegraphics[width=14cm]{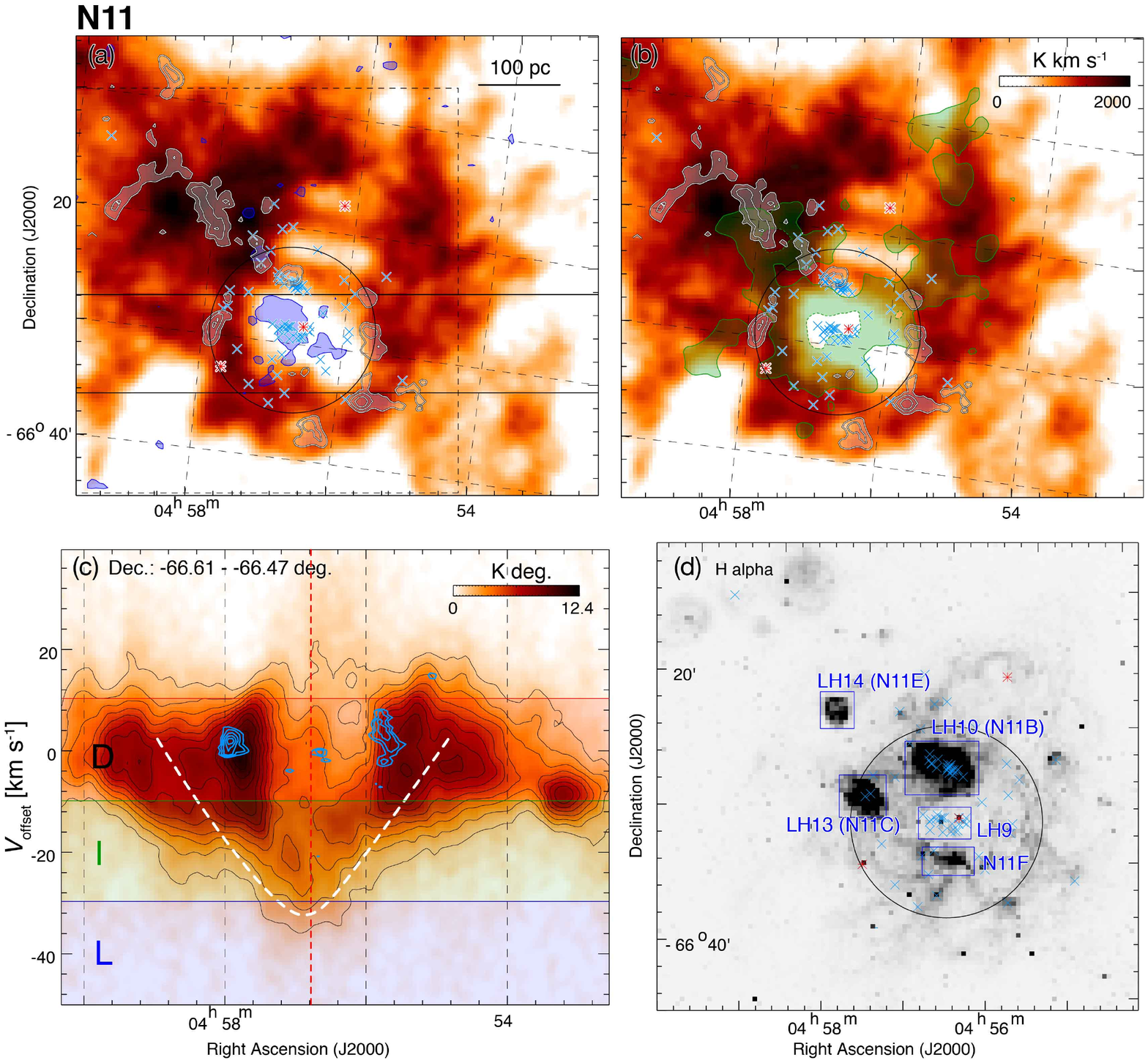}
\end{center}
\caption{ (a) H{\sc i} integrated intensity map of the L-component by blue shaded area ($>$ 150 K km s$^{-1}$) superposed on the D-component by image. (b) H{\sc i} integrated intensity map of the I-component by green shaded area ($>$ 500 K km s$^{-1}$) superposed on the D-component. The white shaded area and contours of (a) and (b) indicate the total integrated intensity map of CO ($>$ 1.5 $\sigma$ (3.89 K km s$^{-1}$); Wong et al. 2011). The contour levels are 3.89, 7.78, and 12.97 K km s$^{-1}$. The red cross, red asterisks, and blue crosses indicate the position of N11 (Henize 1956), WR stars, and O-type stars (Bonanos et al. 2009), respectively. (c) Position--velocity diagram of H{\sc i} by image overlaid with the CO by blue contours. The lowest contour level and interval are 1.5 K deg. and 1.0 K deg. for H{\sc i}; 3 $\sigma$ (0.024 K deg.) and 1 $\sigma$ (0.008 K deg.) for CO. The integration range in the Dec. is from $-$66.61$\degr$ to $-$66.47$\degr$ shown by the black horizontal lines of (a). The red dashed perpendicular line indicates the position of N11 in the R.A.. (d) H$\alpha$ image obtained by the Magellanic Cloud Emission-Line Survey (MCELS; Smith \& MCELS Team 1999) toward N11 shown by black dashed box in (a). The black circle indicates a ring morphology with a cavity of $\sim$100 pc in radius, enclosing OB association LH9 (Lucke \& Hodge 1970). }  
\label{fig9}
\end{figure*}

Detailed spatial distributions of the L-, I-, and D-components around N11 are shown in Figure \ref{fig9}. The weak L-component is located in the cavity of the D-component, and they exhibit complementary spatial distribution as shown in Figure \ref{fig9}(a). The I-component is distributed around the L-component which is connected with the D-component as shown in Figure \ref{fig9}(b). High-mass stars are located toward the L- and I-components, and the distributions of molecular clouds are similar to the I-component. The I-component has ionized holes toward OB associations LH9 and LH14 as shown in Figure \ref{fig9}(d). This trend supports that the I-component is physically associated with high-mass star formation. Figure \ref{fig9}(c) shows the position-velocity diagram of H{\sc i} of the region. Molecular clouds are superposed by contours on H{\sc i}. The integration range in Dec. is from $-$66.61deg. to $-$66.47 deg. The diffuse L-component is seen whose velocity is at $V_{\rm offset}$ $\sim$35 km s$^{-1}$. We found a V-shaped velocity structure in a velocity space (see white dashed line in Figure \ref{fig9}(c) ). Molecular clouds are formed in the velocity range of the D-component where the I- and D-components are overlapped. Figure \ref{fig9}(d) is an H$\alpha$ image \citep[MCELS; ][]{1999IAUS..190...28S} of the star-forming region indicated by a black dashed box in Figure \ref{fig9}(a).

\subsubsection{LHA 120-N79}\label{sec:3.3.3}

\begin{figure*}[htbp]
\begin{center}
\includegraphics[width=\linewidth]{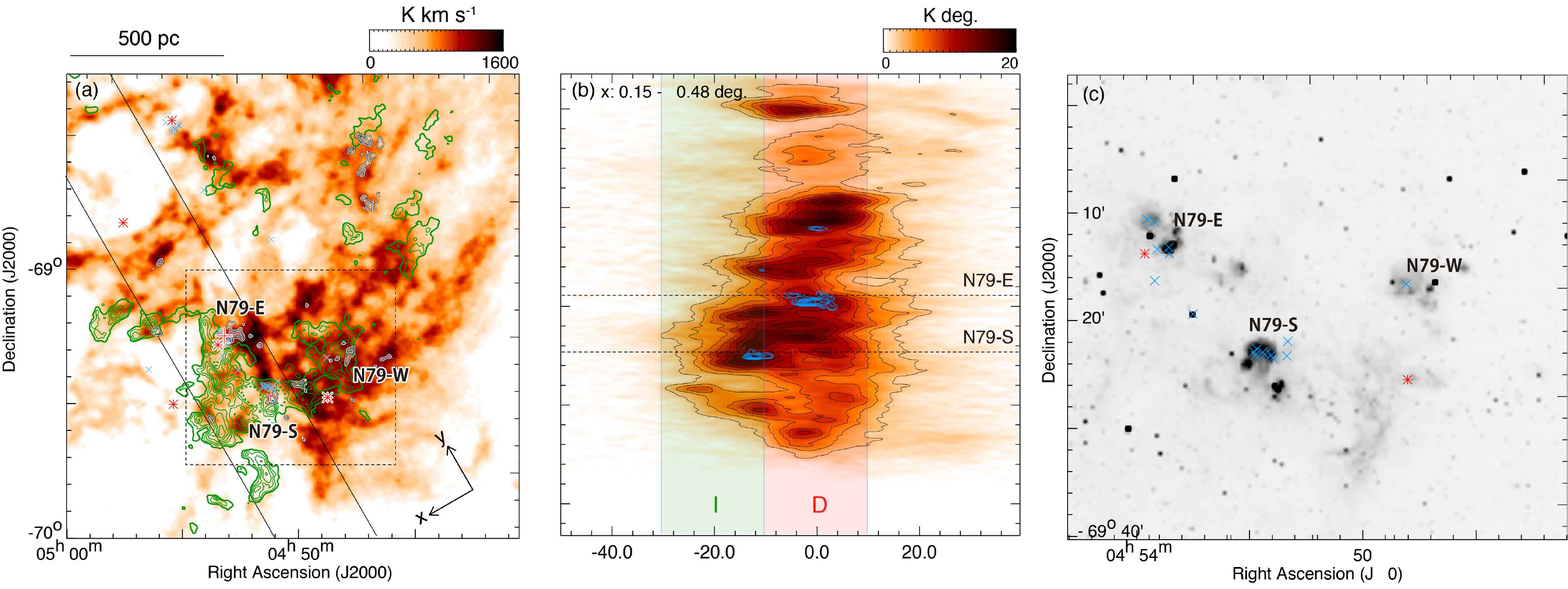}
\end{center}
\vspace*{-3cm}
\caption{ (a) H{\sc i} integrated intensity map of the I-component by green contours superposed on the D-component by image. The lowest contour level and intervals are 350 K km s$^{-1}$ and 100 K km s$^{-1}$. The white shaded and contours indicate the total integrated intensity map of CO ($>$ 1.5 $\sigma$ (3.89 K km s$^{-1}$); Wong et al. 2011). The contour levels are 3.89, 7.78, and 12.97 K km s$^{-1}$. The red cross, red asterisks, and blue crosses of (a) and (c) indicate the position of N11 (Henize 1956), WR stars, and O-type stars (Bonanos et al. 2009), respectively. (b) Position--velocity diagram of H{\sc i} overlaid with the CO contours in blue toward N79. The center of the map at (R.A., Dec.)=(4$^h$51$^m$33.4$^s$, $-$69$^d$01$^m$49.3$^s$) is set as the origin of XY coordinate, and the position angle of the Y axis is $-$30 deg. The integration range in X is from 0.15 deg. to 0.48 deg. The lowest contour level and intervals are 2.5 K deg. and 2.0 K deg. for H{\sc i} and 0.0368 K deg. and 0.0123 K deg. for CO. The positions of N79-E and N79-S are indicated by the dashed horizontal lines. (c) H$\alpha$ image obtained by the Magellanic Cloud Emission-Line Survey (MCELS; Smith \& MCELS Team 1999) toward N79 shown by black dashed box in (a) }  
\label{fig10}
\end{figure*}

Figure \ref{fig10}(a) shows spatial distributions of the I- and D-components around N79. There is few hints of the L-component. The I-component is mainly distributed along the southeastern edge of the D-component. High-mass stars and molecular {are preferentally} located at the boundary of the I- and D-component. We made a position-velocity diagram in the direction of the elongation of the I-component (the black lines of Figure \ref{fig10}(a)) as shown in Figure \ref{fig10}(b). The I-component is connected to the D-component at the positions of N79-E and N79-S regions (see black dashed lines in Figure \ref{fig10}(b)) in a velocity space. Molecular clouds toward N79-S are formed in the velocity range of the I-component. Figure \ref{fig10}(c) is an H$\alpha$ image \citep[MCELS; ][]{1999IAUS..190...28S} of the star-forming region indicated by a black dashed box in Figure \ref{fig10}(a).

\section{Discussion}\label{sec:doscussion}

\subsection{High mass star formation in the I-component}\label{sec:4.1}
The I-component is defined as an intermediate velocity component between the L- and D-components in Paper II. In the present paper, we revealed the spatial distribution of the I-component over the whole LMC for the first time as shown in Figure \ref{fig2}(c). The I component is interpreted as decelerated gas in terms of momentum exchange in the collisional interaction of the L- and D-components as shown by Figures \ref{fig8}(a) and \ref{fig8}(c). If the L- and D-components begin to interact with each other, the I-component can be formed by momentum conservation even if the L-component does not penetrate the D-component. It is interpreted that the difference of column density of the L- and D-components is important to determine velocity structure of H{\sc i} gas after the collision. We suggest that the L-component toward the CO-arc penetrated the D-component with small deceleration by momentum conservation in the eastern part because the column density of the L-component is $\sim$5$\times$10$^{21}$ cm$^{-2}$, which is five times larger than that of the D-component. On the other hand, we presume that the L- and D-components toward the molecular ridge which include R136, N159 etc. (yellow shaded region in Figure 3) are merged into the I-component with significant deceleration probably because they had nearly the same column density. The column density of H{\sc i} gas was elevated to $\sim$1$\times$10$^{22}$ cm$^{-2}$ when the deceleration is strong. As a result, the formation of the high-mass stars R136 etc. were triggered as shown in Figure \ref{fig8}. Thus, the I-component can be interpreted as a good tracer of the collision and high-mass star formation.

Then, we investigated the spatial correlation between the I-component and high-mass stars as shown in section 3.2. From the results of section 3.2, $\sim$74\% of the total number of high-mass stars exhibit spatial correlation between the I-component. Therefore, it is considered that most of the high-mass star formations have been triggered by the collision of H{\sc i} flows that produced the I-component. Figures \ref{fig7}(b), \ref{fig9}(b), and \ref{fig10}(a) show the good spatial correlation with the I-component and high-mass stars/molecular clouds. Around R136, N11, and N79 regions, H{\sc i} gas in the velocity range of the I-component is strongly ionized as seen in Figure \ref{fig3}(c), \ref{fig3}(d), and \ref{fig3}(e). As for R136, only I-component is drastically ionized as seen in a velocity space (Figure \ref{fig8}(a)). These results indicate that the I-component is physically related with high-mass star formation, and are consistent with proposed scenario. In order to confirm this scenario over the whole LMC, we explore evidence for the collision of H{\sc i} flows toward the major star forming regions in section 4.2.
Rest of the high-mass stars ($\sim$26\% of the total number) are mainly located inside of the super giant shells \citep{2013ApJ...763...56D} as shown in Figure \ref{fig19} in Appendix C. It is difficult to investigate the formation mechanism of these high-mass stars because there is little hydrogen gas. 

\subsection{Evidence for the HI Collision between the Two Velocity Components}\label{sec:4.2}

We presented the spatial and velocity structures of the L- ($V_{\rm offset}$:$-$100.1--$-$30.5 km s$^{-1}$), I- ($V_{\rm offset}$: $-$30.5--$-$10.4 km s$^{-1}$), and D- ($-$10.4--9.7 km s$^{-1}$) components by using high-resolution H{\sc i} data in Section 3.3. In Papers I and II, we found observational signatures of the collision of H{\sc i} gas toward R136 and N44. In order to form high-mass stars, it is necessary to achieve a large mass accretion rate of $\sim$10$^{-4}$--10$^{-3}$ $M_{\rm \odot}$ yr$^{-1}$ which makes stellar mass grow against the stellar feedback (Wolfire \& Cassinelli 1987). The mass accretion rate in the shock-compressed layer formed by collision is proportional to the third power of the effective sound speed. The effective sound speed is defined as ($c_{\rm s}^2$+$c_{\rm A}^2$+d$v^2$)$^{1/2}$ , where $c_{\rm s}$ is the sound speed, $c_{\rm A}$ is the Alfven speed, and the d$v$ is the velocity dispersion (turbulence). Therefore, the shock compression by collision amplifies the turbulence and magnetic field, and the mass accretion rate is enhanced enough to form high-mass stars \citep{2013ApJ...774L..31I}. A supersonic velocity separation of dozens of km s$^{-1}$ produces a large mass accretion rate larger than the order of $\sim$10$^{-4}$ $M_{\rm \odot}$ yr$^{-1}$, and makes it possible to form massive cloud cores which is enough to evolve into massive stars.

Most recently, \cite{2018PASJ...70S..53I} conducted the isothermal magnetohydrodynamic (MHD) simulations including self-gravity, and showed that massive filaments with high-mass stars are formed in the shock-compressed layer by a cloud-cloud collision. These simulations indicate that the formation of the filament and the first sink particle (the high-mass protostars) are formed almost at the same time with no significant delay. 
{This mechanism has been investigated by observational studies toward N159W{-South} and N159E{-Papillon} in the south of R136 including the H{\sc i} Ridge region where the L- and D-components are colliding as shown in Paper I. We  calculated molecular mass fraction ($f_{\rm mol}$=$M$(H$_{2}$)/$M$(H{\sc i})) of the H{\sc i} Ridge region. $f_{\rm mol}$ of the L-, I-, and D-components are 20\%, 30\%, and 8\%, respectively. The $f_{\rm mol}$ of the I-component is enhanced and is three times higher than that of the D-component. This suggests that the H{\sc i} gas is effectively compressed by collision and molecular clouds are formed in the I-component.} 
By using the Atacama Large Millimeter/submillimeter Array (ALMA) \citep{2019ApJ...886...14F, 2019ApJ...886...15T}, they found massive filamentary clouds holding high-mass star formation with similar orientation in the {two} regions separated by more than 50 pc. The authors interpreted that the high-mass star formation and the filament formation are not local phenomena within 1--2 pc, but are triggered by a more global event spanning at least 50 pc, and they proposed that the large-scale colliding H{\sc i} flows triggered the coeval formation of {N159W-South and  N159E-Papillon} system. This is consistent with the kpc-scale H{\sc i} collision proposed in Paper I.

The observational signatures of high-mass star formation by colliding H{\sc i} flows are characterized by three elements: (1) the two velocity components with a supersonic velocity separation, (2) the bridge features that connect the two velocity components in velocity space, and (3) the complementary spatial distribution between the two H{\sc i} flows. The molecular clouds and high-mass stars are formed at the shock compressed region by collision. In the present study, we test whether the high-mass star formation in N11 and N79 has been triggered by the colliding H{\sc i} flows by comparing the above signatures with the observational results.

\subsubsection{N11 region}\label{sec:4.2.1}
N11 is one of the largest active star-forming regions in the northwestern corner of the LMC. This H{\sc ii} region was cataloged by \cite{1956ApJS....2..315H}. N11 has a ring morphology with a cavity of $\sim$100 pc in radius, enclosing OB association LH9 (Lucke \& Hodge 1970). {There are} several bright nebulae (N11B, N11C, N11F) around LH9{, whose location is} the center of the cavity. The massive compact cluster HD 32228 dominates this OB association, and has an age of $\sim$3.5 Myr \citep{1999AJ....118.1684W}. LH 10 is the brightest nebula, and lies to the north of LH 9. LH 10 is the youngest OB association with an age of about 1 Myr. Previous studies proposed that star formation of LH10 was possibly triggered by expanding supershell blown by LH9 \citep[e.g., ][]{1996A&A...308..588R, 2003AJ....125.1940B, 2006AJ....132.2653H, 2019A&A...628A..96C}. \cite{1996A&A...308..588R} conducted a detailed analysis on the kinematics of H$\alpha$ emission, and determined an expansion velocity of 45 km s$^{-1}$ for its central hole. The authors suggested that a dynamical age of expansion is 2.5$\times$10$^6$ years, and a possible product of the explosion including three SNe and shock-induced star formation. 
The present study raise{s} another scenario that the I-component plays an important role in high-mass star formation by a statistical investigation (Section 4.1). The I-component is extended to the outside of the H$\alpha$ shell, and there is spatial correlation between the I-component and molecular clouds/high-mass stars not only in the shell-like structure (including N11B, N11F, and N11C) but also in northwest of the shell including N11E as shown in Figure \ref{fig9}(b). Therefore, we propose an alternative scenario of high-mass star formation driven by a large-scale ($>$ 200 pc) collision of H{\sc i} flows.

In N11, the velocity separation of the two components is about 30 km s$^{-1}$ as shown in Figure \ref{fig9}(c). If we consider a projection effect, the actual velocity separation is larger than observed. This value is roughly consistent with the typical velocity separation in the high-mass star forming regions triggered by colliding H{\sc i} flows in the LMC and M33: e.g., R136 ($\sim$50--60 km s$^{-1}$; Paper I), N44 (30--60 km s$^{-1}$; Paper II), and NGC 604 \citep[$\sim$20 km s$^{-1}$; ][]{2018PASJ...70S..52T}. These values are also consistent with the predicted relative velocities of colliding clouds (10--60 km s$^{-1}$) in the LMC shown by \cite{2004ApJ...602..730B}.

We next focus on spatial distributions of the L-, I-, and D-components. In Figure \ref{fig9}(a), the L- and D-components show complementary distribution. The L-component is distributed in a cavity of the D-component. The I-component is distributed around the L-component, and is spatially connected to the L- and D-components as shown in Figure \ref{fig9}(b).
A possible interpretation of these spatial distributions is cloud-cloud collision of H{\sc i} gas. The complementary spatial distribution of the two clouds (or spatial anti-correlation between the two clouds) is one of the significant signatures of collisions because one of the colliding clouds can create an intensity depression in the other cloud. If one of the colliding clouds is smaller than the other, the small cloud creates a cavity of its size in the large cloud as simulated by theoretical studies \citep[e.g., ][]{2014ApJ...792...63T}. The I-component is formed due to the deceleration of the colliding clouds in collision.

Furthermore, we focus on the velocity structures observed in a position velocity diagram. As seen in Figure \ref{fig9}(c), there are I-components between the L- and D-components, which are the bridge features characteris{ed as one of the indications of} cloud-cloud collision. These I-components form a V-shaped structure in the position-velocity diagram as shown by the white dashed line in Figure \ref{fig9}(c). This V-shaped structure provides us with suggestive evidence for head on collision of the L- and D-components. Thus, the I-component is interpreted as a possible tracer of the shock-compressed layer formed by the collision. In addition, most of molecular clouds and high-mass stars are located in the area of the intense I-component (Figure \ref{fig9}(c)).

Age difference of OB associations between LH9 and LH10 at the edge of H$\alpha$ shell could be explained by the difference in the collision epoch. We present a scenario as follows. First, a collision triggered {star formation} at the center of N11, and then the collision proceeded around the shell and formed the younger OB association. It is suspected that the L-component located in the shell penetrated the D-component with small deceleration because of the initial high-density of the L-component. The I-component associated with LH9 has already disappeared by ionization as shown in Figure \ref{fig9}(b). As for the edge of the shell and the northeastern region including N11E, there are only the I- and D-components without L-component. We interpreted that the L- and D-components are merged into the I-component with significant deceleration because they have nearly the same column density as discussed in section 4.1. Moreover, \cite{1992AJ....103.1205P} found that the slope of the initial mass function (IMF) of LH10 is significantly flatter than that of LH9. The final slopes are $\Gamma$= $-$1.6$\pm$0.1 for LH9 and $\Gamma$=$-$1.1$\pm$0.1 for LH 10. It seems that the compression of gas in LH10 is stronger than that in LH9, and it is possible that more high-mass stars formed in LH10. This interpretation is supported by \cite{2019arXiv190908202F} which conducted a detailed analysis of the results of the numerical simulations of magneto-hydrodynamics which deal with colliding flows at a relative velocity of 20 km s$^{-1}$ \citep{2013ApJ...774L..31I}. They found that the mass function of the dense
cores is significantly top-heavy (flatter) as compared with the universal IMF, indicating that the cloud-cloud collision triggers preferentially the formation of O and early B stars. These signatures are consistent with the possible scenario that the molecular clouds and high-mass stars in N11 were formed by collision between the L- and D-components.

\subsubsection{N79 region}\label{sec:4.2.2}
N79 is a H{\sc ii} region cataloged by \cite{1956ApJS....2..315H} and located in the southwestern corner of the LMC. Figure \ref{fig10}(c) shows H$\alpha$ image toward N79. The N79 region attracts attention as an earliest stage of active star formation. N79 harbors three Giant molecular clouds (GMCs) such as N79-E, N79-S, and N79-W as shown in Figure \ref{fig10}(a). \cite{2017NatAs...1..784O} focused on the most luminous YSO (H72.97-69.39) in the LMC. The authors suggested that this object might be an SSC candidate because of its high star formation efficiency. Subsequently, \cite{2019ApJ...877..135N} revealed filamentary CO clouds associated with the SSC candidate. They also proposed that the location of 30 Doradus including R136 and N79 that the intersection of the tidal arms and stellar bar-ends possibly ideal physical conditions to create SSC. 
Alternatively, there is a possibility that other factors may play a role in active star formation such as accretion flows \citep{2015Natur.519..331T} or tidal interactions \citep{2007PASA...24...21B}. In the present paper, we will discuss an alternative scenario that high-mass star formation is triggered by tidally-driven colliding H{\sc i} flows. Particularly, H{\sc i}/CO gas are concentrated in the southeast (the H{\sc i} Ridge region) of the LMC as shown in Figure \ref{fig1}. It is warranted to explore if the tidally-driven gas dynamics induces the star formation.

In N79 region, the D- and I-components are observed, and the velocity difference is $\sim$20 km s$^{-1}$ as shown in Figure \ref{fig10}(b). This value is larger than the typical velocity dispersion of the D-component as shown in Figure \ref{fig4}(b), and also consistent with previous studies \citep[Paper I; Paper II; ][]{2018PASJ...70S..52T}. A possible interpretation is that the I- and D-components collided at velocity of 20 km s$^{-1}$. The relative velocities of colliding clouds predicted theoretically by Bekki (2004) are in a range from 10 to 60 km s$^{-1}$, and consistent with observed velocity difference. In Figure \ref{fig10}(a), the I- and D-components show spatial anti-correlation, a characteristic signature of collision between the two components. The I-component exists along the edge of the D-component. The I-component (small cloud) does not form a cavity in the D-component (large cloud). 
We next focus on the velocity distribution. We made a position--velocity diagram along the elongation of the I-component as shown in Figure \ref{fig10}(c). The I- and D-components are connected in a velocity space, whereas a V-shaped distribution similar to N11 is not found. These results are interpreted as collision between the I-component and the southeastern edge of the D-component. This off-center collision has been observed in the Galactic H{\sc ii} region of NGC 2068/2071 without V-shaped velocity structure by \citet{2020PASJ..tmp..163F}, which also showed new simulations of an off-center collision. The spatial distribution and velocity structure of N79 look qualitatively similar to results of NGC 2068/2017. 
Finally, we compared the spatial distributions of the molecular clouds, the high-mass stars, and the I-component. Most of the molecular clouds and the high-mass stars are distributed along the boundary of the I- and D-components. If we assume that the collision is triggered at the southeastern edge of the D-component, it is likely that the molecular clouds and the high-mass stars were formed in the compressed layer of the collision.

We are able to analyze high-spatial resolution data of molecular clouds in order to test the proposed scenario. We will execute observations of filamentary clouds toward N44, N11, and N79 by using ALMA (2019.2.00072.S) is in progress. For one of the molecular clouds of N79, filamentary molecular clouds have already found with ALMA \citep{2019ApJ...877..135N}, which is consistent with a scenario that filamentary clouds were formed by the collision. We also have to analyze H{\sc i} gas at a resolution comparable to ALMA observations in order to better understand the physical connection between filaments and large-scale H{\sc i} flows.

The origin of the H{\sc i} collision is possibly tidal interaction between the LMC and SMC. The L-component is metal poor toward the H{\sc i} Ridge and N44 in the Diffuse L-component as shown in Paper I/II. \cite{2007MNRAS.381L..16B} also found that the metal poor gas from the SMC has continued to flow in the LMC since 0.2 Gyr ago \citep[see also Figure 1 of ][]{2007PASA...24...21B}. We will show detailed Metallicity map over the whole LMC elsewhere in a separate paper.

\begin{figure}[htbp]
\begin{center}
\includegraphics[width=\linewidth]{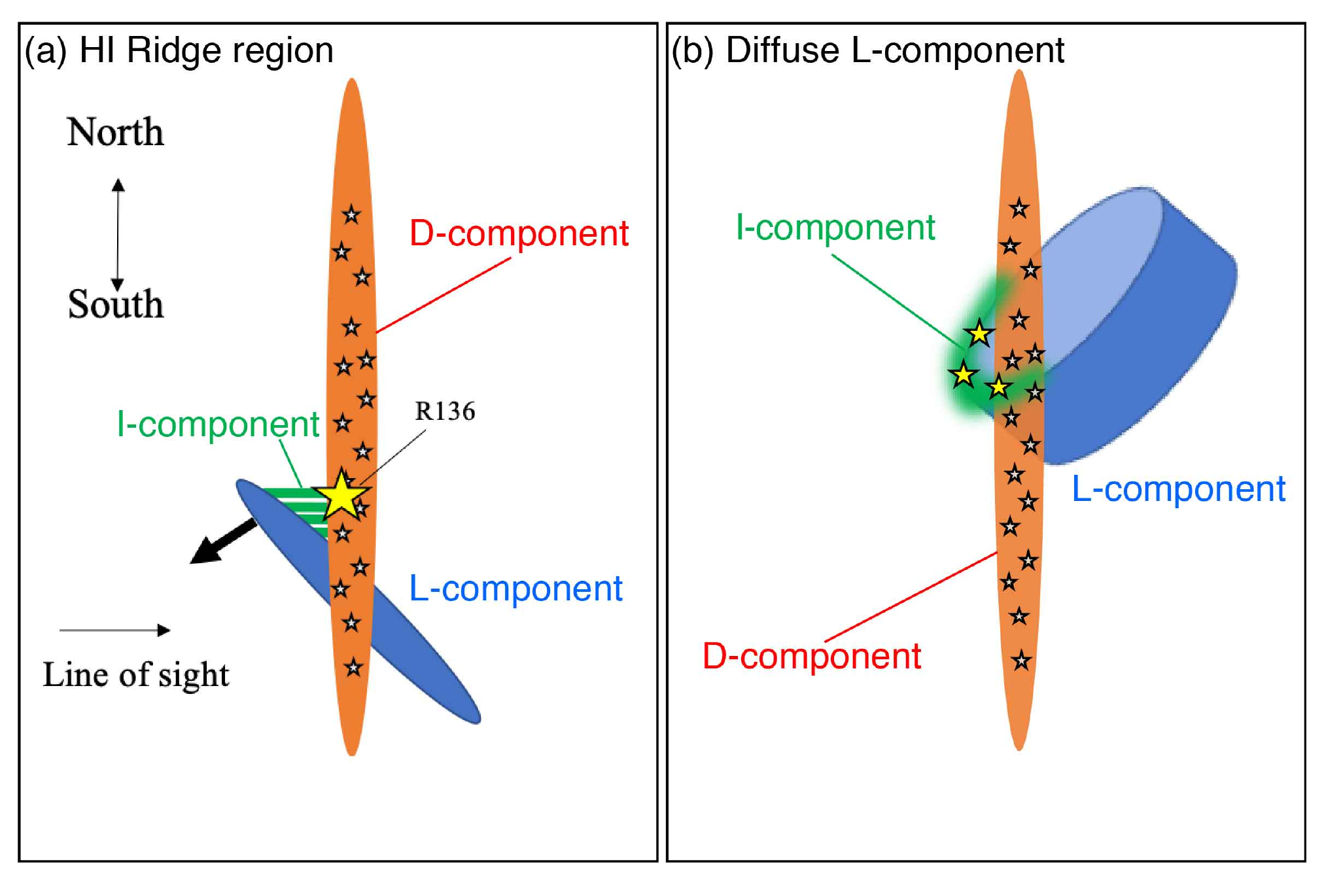}
\end{center}
\caption{Schematic views of the collision scenario between the Diffuse L-component and the D-component during the collision (a). Orange and blue disks are the D- and L-component, respectively. The green image indicates the I-component produced by the interaction between the L- and D-components. }  
\label{fig11}
\end{figure}%

\subsection{Geometry of the HI collision}\label{sec:4.3}
From Sections 4.1 and 4.2, the molecular clouds and $\sim$70\% of the high-mass stars were possibly formed by large-scale colliding H{\sc i} flows. In addition, it is found that the high-mass star formation was not triggered unless the I-component is formed by collision even if the L- and D-components exist. For example, in most of the Diffuse L-component without the I-component, there is almost no high-mass star formation. On the other hand, the I-component exist in the southern part of the H{\sc i} Ridge, but high-mass star formation is not seen as shown in Figure \ref{fig5}(a). We will pursue the cause for this trend by considering the three-dimensional structure of the collision investigated by previous studies in the following.  Detailed investigation by the comparison with numerical simulations are shown in Section 4.4.

\subsubsection{HI Ridge region}\label{sec:4.3.1}
For the H{\sc i} Ridge region, it is proposed that the collision of the L- and D-components proceed to the south from the north by the comparison of the Av map and total hydrogen column density map \citep{2019PASJ...71...95F}. The authors compared the Av map and the L-, I-, and D-components, and found that the L-component is located in front of the D-component at north of $\delta$J2000.0 =$-$70.8$\pm$0.2 at $\alpha$J2000.0 = 87.4 deg. This result supports the scenario that the collision is on-going and the southern part of the H{\sc i} Ridge is before/beginning of the collision as shown in Figure 11a \citep[see also schematic view of the geometry of H{\sc i} gas as shown in Figure 7 of ][]{2019PASJ...71...95F}. This geometry well explains the activity of high-mass star formation.  Around Dec. = $-$68 deg. to $-$70 deg., the high-mass star formation is active, but at Dec. = $-$70 deg. to $-$71 deg., there is only the I-component without active high-mass star formation. Therefore, the I-component which does not make high-mass stars in the H{\sc i} Ridge region is at an early stage of the collision, and it is thought that high-mass star formation will begin in future. 

\subsubsection{Diffuse L-component}\label{sec:4.3.2}
For the Diffuse L-component, there are also high-mass star formation triggered by colliding H{\sc i} flows induced by the tidal interaction. In Paper II, the observational signatures of colliding H{\sc i} flows toward N44 are found in the Diffuse L-component. The authors also found evidence for less dust abundance in the N44 region. This is ascribed to the tidal interaction between the LMC and the SMC 0.2 Gyr ago, and the inflow of metal poor gas from the SMC is probably responsible for less dust abundance. Figures \ref{fig20} (j) (k) of Appendix D show that the L- and D-components are connected in a velocity space. Thus, the L- and D-components are colliding over the whole Diffuse L-component.

We also investigated geometry of the collision based on the H{\sc i} data such as the 1st moment map, distribution of the I-component, and column density of the L- and D-components. As shown in Figure \ref{fig2}(a), some regions (nearby N105 and N119) in the southern part of the Diffuse L-component show higher H{\sc i} intensity than in the northern part.  Maximum column density of the L-component is $\sim$2.6$\times$10$^{21}$ cm$^{-2}$ whereas the D-component has a column density of 3.2$\times$10$^{20}$ cm$^{-2}$, which is an order of magnitude smaller than that of the L-component at the same position. Therefore, it is interpreted that the L-component penetrated the D-component in the same way as a case of the H{\sc i} Ridge which is discussed in Section 4.1. Moreover, the 1st moment map shows that the L-component is decelerated at the southern edge as shown in Figure \ref{fig4}(a). The I-component is also distributed along the southern edge of the L-component as shown in the white dashed line of Figure \ref{fig3}. So, it is possible that the L- and D-components are merging to form the I-component in the south of the Diffuse L-component.

On the other hand, the northern part of the L-component is not decelerated and the 1st moment is $\sim$$-$50 to $-$60 km s$^{-1}$. There is no I-component toward this region as shown in Figure \ref{fig3}, so the L-component is possibly before or in the beginning of collision. Moreover, there are no significant molecular cloud/high-mass star formation, which is consistent with that no significant compression by collision is taking place.  We therefore propose that the collision possibly proceeds to the north from the south as shown in Figure \ref{fig11}b. It is desirable that this geometry is investigated by comparison with extinction map and numerical simulation. 

\begin{figure*}[htbp]
\begin{center}
\includegraphics[width=\linewidth]{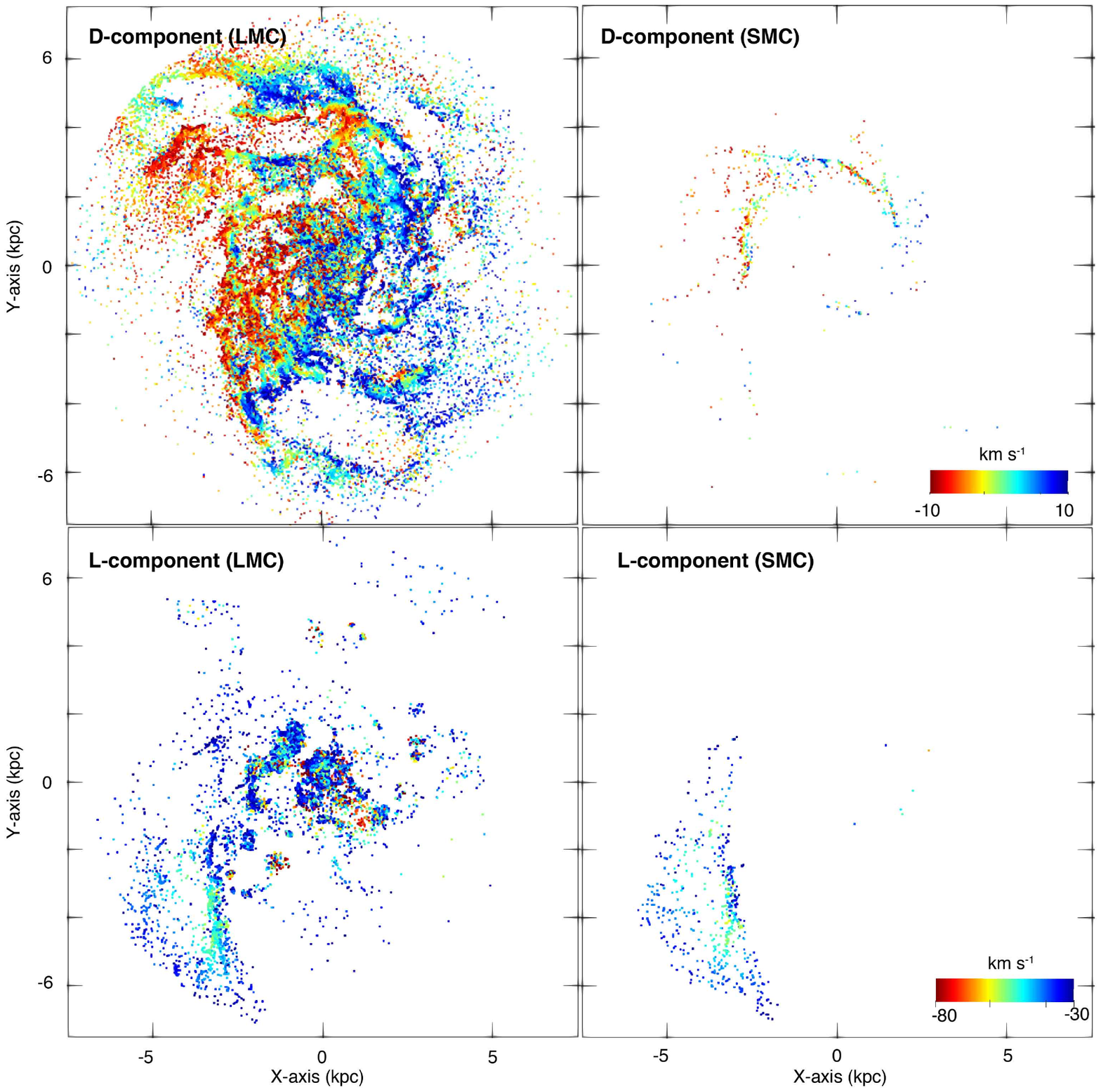}
\end{center}
\caption{Spatial distributions of the LMC (left) and SMC gas particles (right) in the D-component with $-$10$<$$V$ $<$ $-$10 km s$^{-1}$ (upper), where $V$ is the line-of-sight velocities of gas particles, and in the L-component with $-$100$<$ $V$ $<$ $-$30 km s$^{-1}$ (lower) for the best LMC in the present study. The simulated LMC disk is rotated so that the strong gaseous arm can be placed in the lower left panel of this x-y projection as observed. Colors show the line-of-sight velocities of gas particles.}  
\label{fig12}
\end{figure*}%

\subsection{Comparison with numerical simulations}\label{sec:4.4}
We carried out new numerical simulations of the interaction between the LMC and SMC in order to confirm our collision picture from a theoretical point of view. The detailed setup of the simulations is given in the Appendix E.

The numerical simulations have successfully reproduced the detailed distribution of the L-component and its 3D trajectory. We find that the L-component is moving toward us from the farside of the D component in the projected direction on the LMC disk from the northwest to the southeast. The projected direction of the motion is consistent with the collision path of the L component derived before by Fukui et al. (2017) and also by the highly directive filamentary clouds observed with ALMA in N159W{-South} and N159{E-Papillon} (Fukui et al. 2015; 2019; Saigo et al. 2017; Tokuda et al. 2019) which are suggested to have been formed by the same kpc-scale collision.

Fig. 12 shows the spatial distributions of gas particles originating from the LMC and the SMC in the D- and L-components. In order to make a same consistent compari- son between observations and simulations, we here apply the same velocity ranges observed for the two components to the gas particles of the simulations. This Fig. 12 can be compared with the observed distributions of the two components, be- cause the simulated LMC is properly rotated to match the observed locations of the spiral arms of the LMC. It is clear in this Fig. 12 that the gas particles in the L-component are located preferentially in (i) the arm in the lower left part of the LMC gas disk, (ii) the central barred region, and (iii) the upper part of the LMC disk. These spatial distributions of the L-component are broadly consistent with observations, which suggests that the LMC-SMC direct collision is respon- sible for the formation of the L-component. Intriguingly, the L-component consists of both LMC and SMC gas particles, and even the D-component contain{s some of the SMC gas, which} have been only recently accreeted onto the LMC disk, not during the last LMC-SMC collision about 0.2 Gyr ago.

In this model, the total mass of the L-component is only 12\% of the D-component, though the mass-ratio of the two depends on the details of the present dynamical models (in particular, the orbits of the LMC and the SMC and the initial gas mass of the SMC). This recent gas accretion of the SMC gas disk and the resultant formation of the L- component has been first demonstrated to be possible in the present study. SMC gas particles accreted earlier onto the LMC can lose kinetic energy due to the dissipative collision with the LMC gas disk so that they can settle down onto the D-component disk with lower line-of-sight velocities at the present time. This earlier accretion might have triggered the formation of massive stars in the upper (i.e., northern) part of the LMC disk. 

\begin{figure*}[htbp]
\begin{center}
\includegraphics[width=15cm]{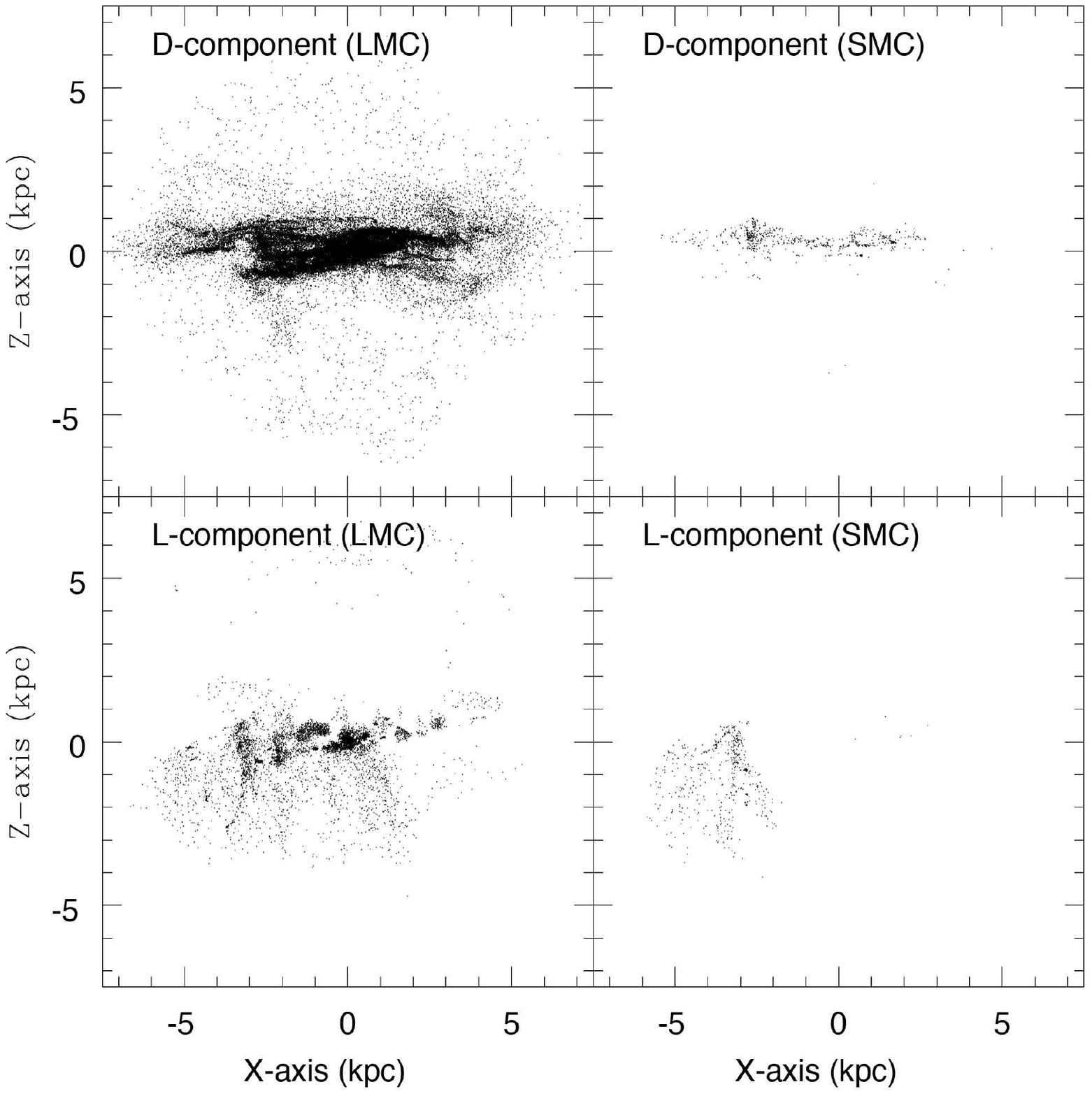}
\end{center}
\vspace*{-5cm}
\caption{The same as Fig. 1 but for the edge-on view of the LMC disk.}  
\label{fig13}
\end{figure*}%

Fig. 13 clearly shows that the L-component consists mostly of gas either within the LMC disk ($\|z\|$ $<$ 1 kpc) or those behind the LMC disk (i.e., z $<$ $-$1 kpc). The LMC gas particles in the L-component were pushed out once through hydrodynamical interaction between the gas disks of the LMC and the SMC. Also, a minor fraction of the LMC gas particles were expelled from the disk through supernova feedback effects to become the L-component. Intriguingly, a very minor fraction of the LMC gas particles in the D- component can be seen well above the LMC disk (z $>$ 1 kpc). Furthermore, the D-component contains the LMC gas particles well above and behind the LMC disk ($\|z\|$ $>$ 1 kpc). These “halo” gas in the D-component will become the L- component soon after they start falling onto the LMC disk with high velocities. This implies that if high-velocity gas cloud collisions are responsible for the massive OB star formation, as proposed in our previous papers (e.g., Fukui et al. 2017), then such star formation will be able to continue due to rapid infall of this halo gas.

\begin{deluxetable}{ccc}
\tablewidth{7.0cm}
\tablecaption{Statistical properties of cloud-cloud collisions in the LMC. Here cloud-cloud collisions with relative velocities larger than 50 km s$^{-1}$ are referred to as “high-velocity”. The details of the method to find cloud-cloud collisions in the LMC are given in the main text.}
\label{tab:2}
\tablehead{\multicolumn{1}{c}{Physical properties } &Values }
\startdata
Number fraction of colliding clouds& 0.098  \\
Number fraction of high-velocity collisions& 0.009 \\ 
Mean collisional velocity & 11.6 km s$^{-1}$\\ 
Maximum collisional velocity& 107.9 km s$^{-1}$\\ 
\enddata
\end{deluxetable}

\begin{figure*}[htbp]
\begin{center}
\includegraphics[width=\linewidth]{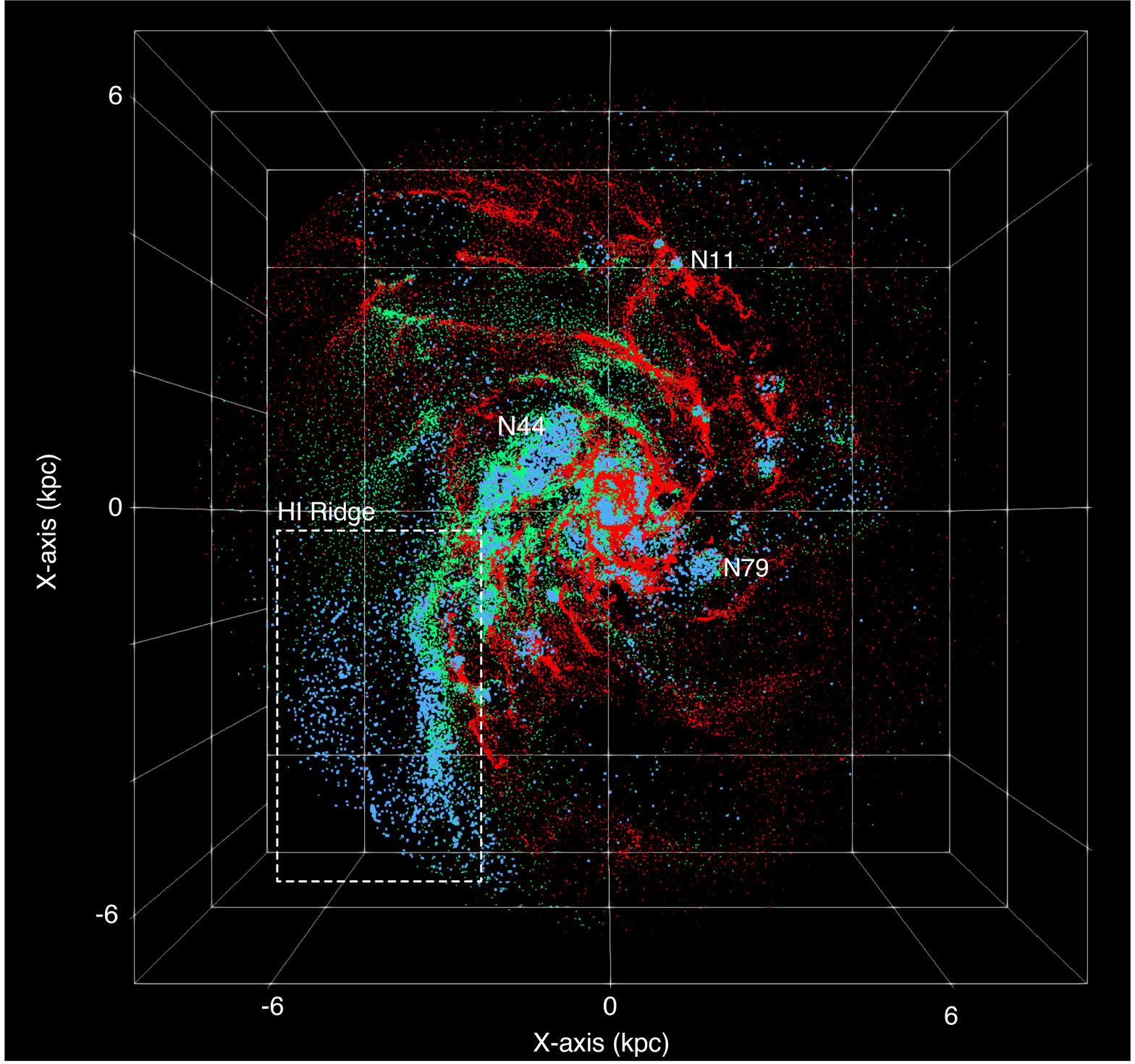}
\end{center}
\caption{Spatial distributions of the gas particles. Red, green, and light blue show the D-, I-, and L-components, respectively. }  
\label{fig14}
\end{figure*}%

Fig. 14 shows that the numerical result reproduced the spatial distributions of three velocity components which are corresponding to the L-, I-, and D-components of the observational result as shown in Fig.3. In the HI Ridge region as shown by white dashed box, gas particles reproduced arcuate structures that consisted of the L-component and CO-arc and straight structure that consisted of I-component and molecular ridge. Toward N44, N79, and N11 we found the L- and I-components. These results support the scenario that collisions of H{\sc i} flow induced by tidal interaction was triggered over the whole LMC.

\begin{figure*}[htbp]
\begin{center}
\includegraphics[width=\linewidth,clip]{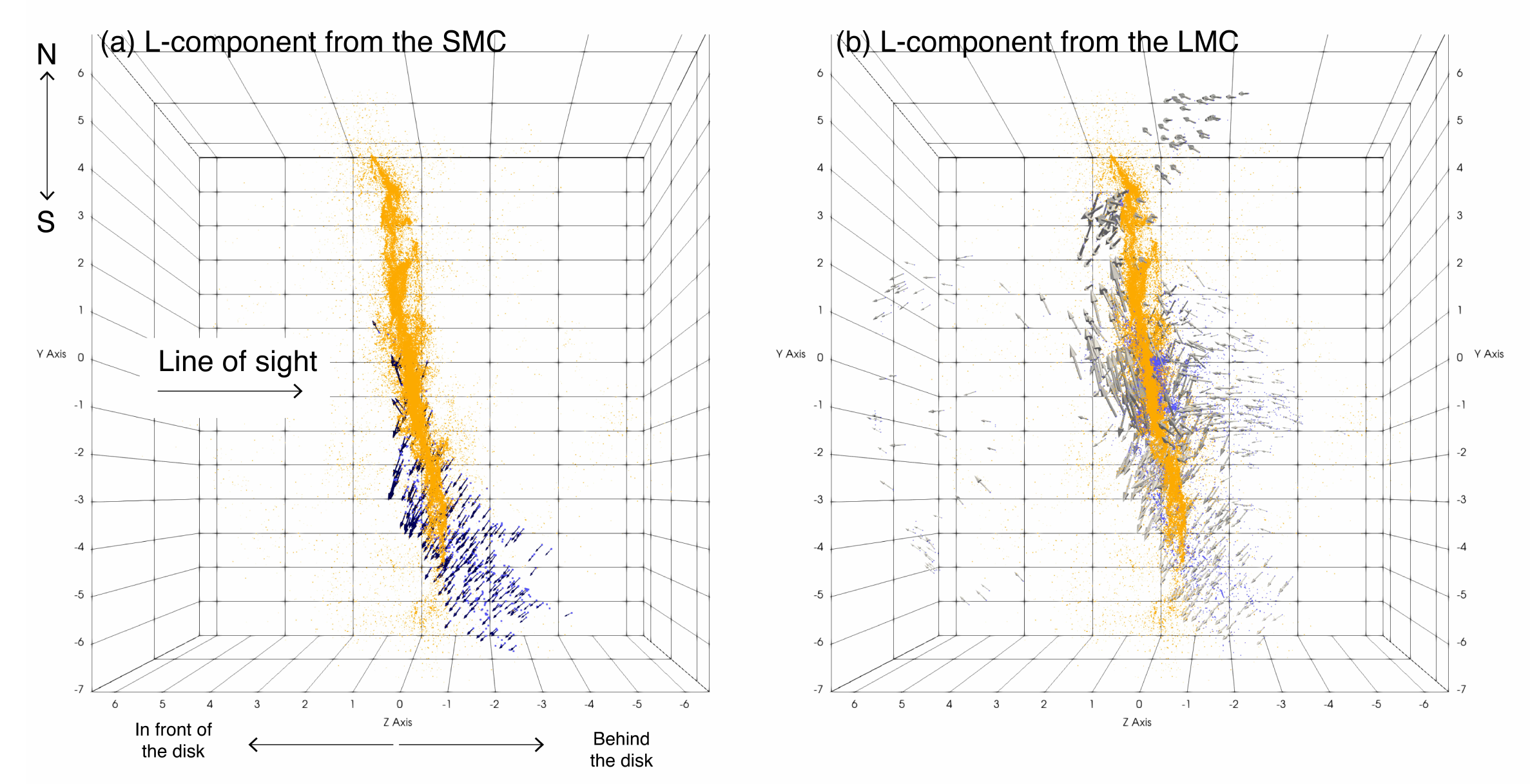}
\end{center}
\caption{Geometry of the collision between the L- and D-components estimated by the numerical results in z-y space.(a) Orange and blue particles indicate all gas particles in the D-components and the SMC gas particles in the L-component, respectively. Arrows show the velocity vectors of particles in the L-component. (b) Same as (a), but blue particles show the LMC gas in the L-component.}  
\label{fig15}
\end{figure*}%

Figure15 shows geometry of the collision between the L- and D-components estimated by the numerical results. We derived 3-dimensional velocity vectors of particles of the L-component. We plotted the spatial distributions of the L- and D-components in z-y space. Figure 15a shows the motion of the SMC gas particles in the L-component from. Most of the SMC gas particles are distributed toward the H{\sc i} Ridge region as shown in Figure 12. The L-component has tilt and particles consists of the L-component are moving toward lower left from the upper right. In north part of the H{\sc i}  Ridge, the L-component penetrated the D-component and located in front of the D-component. In southern part of the H{\sc i} Ridge, the L-component is still behind the D-component and will collide with D-component in the future. This 3-dimensional geometry of the collision is consistent with schematic view of the collision shown in Figure 11a. 

Figure 15b shows the motion of the L-component from the LMC. The LMC gas in the L-component. The motion of the LMC gas particles are very complicated because these particles were pushed out once through hydrodynamical interaction between the gas disk of the LMC and the SMC. There are many particles before collision around the center of the disk, and it is possible that the collision that shown in Figure 11b is taking place.

In order to confirm that high-velocity (V $>$ 50 km s$^{-1}$) cloud-cloud collision required for massive OB star formation is possible in the LMC gas disk, we have investigated the relative velocities of all pairs of two (SPH) gas particles with the mutual distances less than 100 pc. About 10\% of the LMC gas particles are found to be currently colliding with the SMC gas particles within the LMC disk. This high fraction of colliding LMC gas particle implies that the LMC is still strongly disturbed by the last LMC-SMC interaction and thus turbulent. The number fraction of these colliding LMC gas that has V $>$ 50 km s$^{-1}$ is only 9\%, which suggests that high-velocity cloud-cloud collision cannot be a major mode of star formation (Table 2). However, the presence of such high-velocity gas collision in the simulated LMC strongly sup- ports the scenario that the LMC-SMC collision can finally trigger the massive OB star formation due to high-velocity cloud-cloud collision.

Thus, these results demonstrate that both the formation of the L-component and the formation of colliding gas clouds with high velocities wit $V$ $>$ 50 km s$^{-1}$ are due to the last LMC-SMC collision with a small pericenter distance ($R$ $<$ 5 kpc). It should be stressed here that the accretion event of the SMC gas onto the LMC can occur quite recently, though the LMC-SMC collision was about 0.2 Gyr ago. This is mainly because it takes more than 100 Myr for the SMC gas to be transferred from the SMC to the LMC after the last LMC-SMC collision. It is also confirmed that a minor fraction of the LMC gas can be transferred to the outer edge of the SMC and the Magellanic bridge region to collide with the SMC gas. Although these collisions in the SMC and the bridge regions are quite important in under- standing the formation of massive stars in these regions, we will discuss these issues in our forthcoming papers. 

\section{Conclusions}\label{Conclusions}
{We complehensively analyzed the H{\sc i} data across the LMC and the SMC.} The main conclusions of the present paper are summarized as follows. 

\begin{enumerate}
\item   We analyzed the high angular resolution H{\sc i} data (60$\farcs$ corresponding to $\sim$15 pc at the distance of the LMC) observed by the ATCA and Parkes telescopes \citep{2003ApJS..148..473K}. The spatial distribution of the I-component {was} revealed over the whole LMC for the first time. The I-component is the intermediate velocity component between the two HI{\sc i} components with velocity difference of $\sim$60 km s$^{-1}$ (the L- and D-components) produced by deceleration of the colliding clouds. 

\item The distribution of the I-component exhibit{ed} spatial correlation between the high-mass stars over the whole LMC. $\sim$70 \% of the high-mass star are associated with the I-component whose integrated intensity is larger than 300 K km s$^{-1}$. This trend {was} significantly different from what is expected in the case of a purely random distribution. The result suggest{ed gas compressions by H{\sc i} collisions made it possible to form molecular clouds and high-mass stars.}

\item  We revealed the spatial and velocity distributions of H{\sc i} Ridge, N11, and N79. We confirm{ed} that two H{\sc i} components at different velocities {were} colliding at 500 pc scale with a velocity difference of $\sim$30--60 km s$^{-1}$. The collision {was} characterized by the spatial complementary distribution (anti-correlated distribution) and bridge features in a velocity space.

\item We also investigate{d} the geometry of the collision. it {wa}s shown that the colliding H{\sc i} gas, which {was} probably plane-like with some tilt to the D-component. The tilt naturally explain{ed} the different epoch of triggering from place to place, thereby offering an explanation on the cause of the age difference among OB associations in the H{\sc i} Ridge etc.

\item New numerical simulations successfully reproduced the detailed distribution of the L-component and its 3D trajectory. We {found} that the L-component {was} moving toward us from the farside of the D-component in the projected direction from the northwest to the southeast. The projected direction orbit is consistent with the collision path of the L component derived the complementary distribution between the L- and D-components by Fukui et al. (2017) and also by the highly directive filamentary clouds observed with ALMA in N159W-South and N159E-`apillon (Fukui et al. 2015; 2019; Saigo et al. 2017; Tokuda et al. 2019) which are suggested to have been driven by the same kpc-scale collision.

\end{enumerate}

\acknowledgments
The NANTEN project is based on a mutual agreement between Nagoya University and the Carnegie Institution of Washington (CIW). We greatly appreciate the hospitality of all the staff members of the Las Campanas Observatory of CIW. We are thankful to many Japanese public donors and companies who contributed to the realization of the project. This study was financially supported by JSPS KAKENHI Grant Number 15H05694. This work was also financially supported by Career Development Project for Researchers of Allied Universities. The ATCA, Parkes, and Mopra radio telescope are part of the ATNF which is funded by the Australian Government for operation as a National Facility managed by CSIRO. The UNSW Digital Filter Bank used for the observations with the Mopra Telescope was provided with support from the Australian Research Council. Based on observations obtained with Planck, an ESA science mission with instruments and contributions directly funded by ESA Member States, NASA, and Canada. The Southern H-Alpha Sky Survey Atlas, which is supported by the National Science Foundation.
Cerro Tololo Inter-American Observatory (CTIO) is operated by the Association of Universities for Research in Astronomy Inc. (AURA), under a cooperative agreement with the National Science Foundation (NSF) as part of the National Optical Astronomy Observatories (NOAO).  The MCELS is funded through the support of the Dean B. McLaughlin fund at the University of Michigan and through NSF grant 9540747.

\begin{figure*}[htbp]
\begin{center}
\includegraphics[width=12cm]{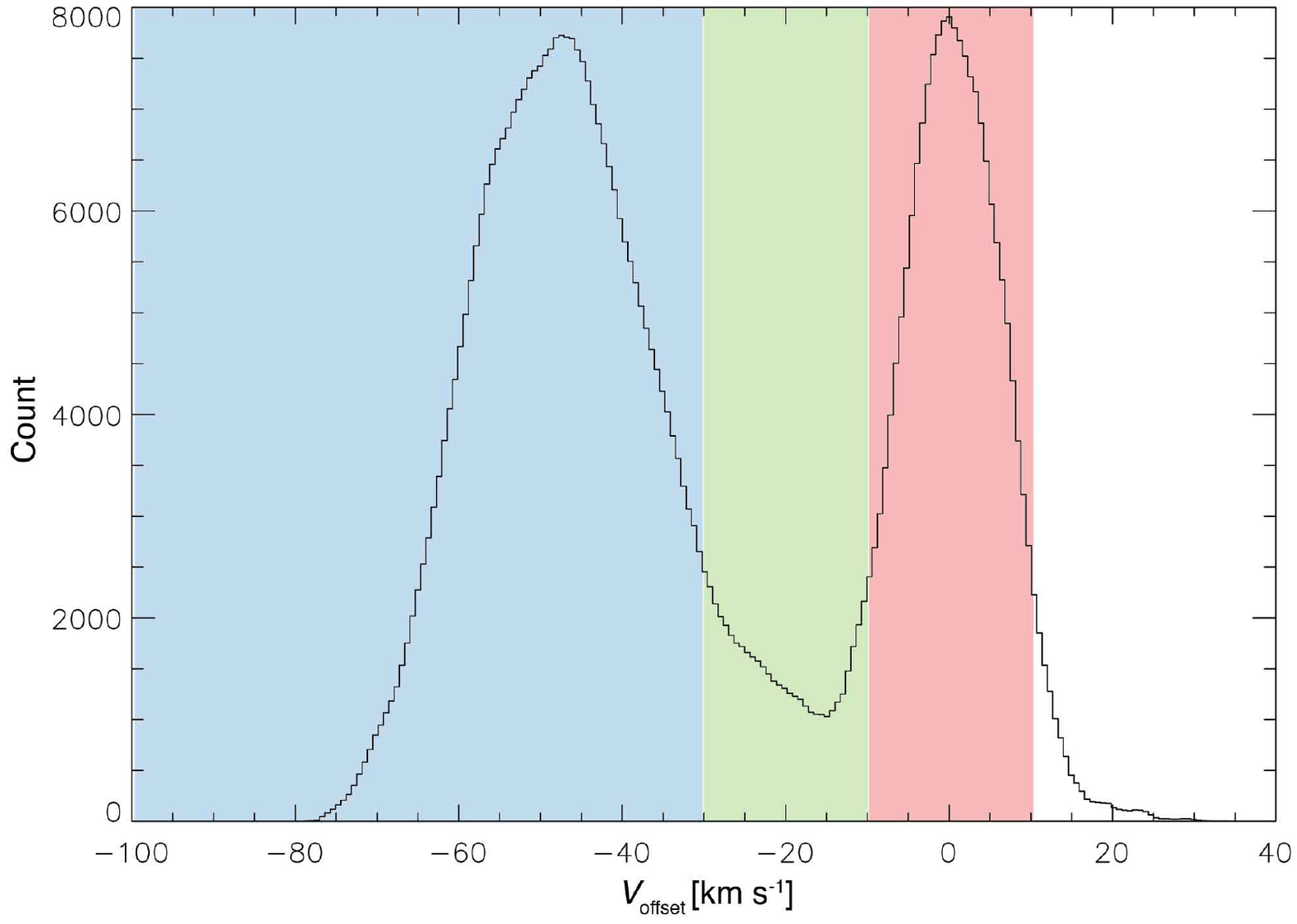}
\end{center}
\vspace*{-5cm}
\caption{{Histogram of the number of pixel whose brightness temperature of H{\sc i} is greater than 30 K toward the northern part of H{\sc i} Ridge (R.A.=86.045975 deg.--88.766197 deg., Dec.=-69.122518 deg.---70.003540 deg.). The horizontal and vertical axis are $V_{\rm offset}$ and the number of pixel within the each $V_{\rm offset}$ bins, respectively. The blue, green, and red show the velocity ranges of the L-, I-, and D-components, respectively.}}  
\label{fig16}
\end{figure*}

\appendix
{
\section{Velocity ranges of the L-, I-, and D-components}
Figure \ref{fig16} shows Histogram of the number of pixel whose brightness temperature of H{\sc i} is greater than 30 K toward the northern part of H{\sc i} Ridge (R.A.=86.045975 deg.--88.766197 deg., Dec.=-69.122518 deg.---70.003540 deg.). There are three velocity components in the histogram and they are corresponding to the L-, I-, and D-components.}

\section{Velocity channel maps of H{\sc i} toward R136, N11, and N79}
Figure \ref{fig17}, \ref{fig18}, and \ref{fig19} are velocity channel maps of H{\sc i} toward R136, N11, and N79, respectively. The  velocity ranges from $-$48.1 to 39.6 km s$^{-1}$, with an interval of 3.25 km s$^{-1}$.

\section{Distributions of high-mass stars correlated with the I-component}
Figure \ref{fig20} shows the distributions of high-mass stars correlated with the I-component in (a) and not correlated with the I-component in (b). 

\section{Channel maps of the position velocity diagrams over the whole LMC}
We show the 11 right ascension--velocity diagrams of H{\sc i} over the whole LMC in Figure \ref{fig21}. The integration range is 0.27 deg. ($\sim$235.6 pc), and the integration range is shifted from north to south in 0.27 deg. step.

\section{Hydrodynamical simulations of the LMC and the SMC}
Using the original hydrodynamical simulations of the LMC and the SMC by Bekki \& Chiba (2007a) and Yozin \& Bekki (2014), we here investigate the gas transfer between the LMC and the SMC over the last 0.3 Gyr (i.e., during and after the LMC-SMC collision) in detail. The main purpose of this investigation is to confirm that the proposed collision of the LMC gas disk with cold gas originating from the SMC gas disk is really possible. We further investigate whether the observed L- and D-components of the LMC gas can be formed during the last LMC-SMC collision. Since the details of the simulation code are already given in Bekki (2013) and Yozin \& Bekki (2014), we here briefly describe the code in the present study. Although we have run $\sim$20 models, we only present only one model that can reproduce the observed D- and L-components of the LMC very well in the present study. 

\subsection{The model}
We numerically investigate the dynamical and hydrodynamical evolution of interacting gas-rich dwarf galaxies, the LMC and the SMC, over the last 0.3 Gyr using our original chemodynamical simulation code with dust physics. The new code (Bekki 2013, 2015) can be run on GPU machines (clusters) and it adopts the smoothed-particle hydrodynamics (SPH) method for investigating temporal spatial variations of gas in galaxies and star-forming gas clouds. Each of the two dwarfs is assumed to be composed of dark matter halo, stellar disk, and gaseous disk in the present study: no bulge is assumed. The total masses of dark matter halo, stellar disk, and gas disk in the LMC (SMC) are denoted as $M_{\rm h,l}$ ($M_{\rm h,s}$) $M_{\rm s,l}$ ($M_{\rm s,s}$) and $M_{\rm g,l}$ ($M_{\rm g,s}$) respectively. We adopt the density distribution of the NFW halo (Navarro, Frenk \& White 1996) suggested from CDM simulations and the “$c$-parameter” ($c$ = $r_{\rm vir}$/$r_{\rm s}$, where $r_{\rm vir}$ and $r_{\rm s}$ are the virial radius of a dark matter halo and the scale length of the halo) and $r_{\rm vir}$ are chosen appropriately for a given dark halo mass ($M_{\rm h}$) and the $c$ parameter is set to be 12 for the two dwarfs.

The radial ($R$) and vertical ($Z$) density profiles of the stellar disk of the LMC are proportional to exp($-R$/$R_{0}$) with scale length $R_{0}$ = 0.2 $R_{\rm s}$ and to sech$^2$($Z$/$Z_{0}$) with scale length $Z_{0}$ = 0.04 $R_{\rm s}$, respectively. This exponential disk with different scale length and hight is adopted for the SMC. The gas disk size $R_{\rm g}$ is assumed to be the stellar disk size ($R_{\rm s}$) both for the LMC and the SMC, though tidal and ram pressure stripping of gas and stars from the LMC and the SMC can cause the different sizes in the two components. Star formation and chemical evolution are both properly modeled using our previous models for these processes (Yozin \& Bekki 2014). Since the LMC has a stellar bar, we first run the isolated LMC model for 1 Gyr to form a stellar bar in the disk and then use it as an initial model for the LMC in the present study. The basic parameters for dark matter halos, stellar disks, and gaseous disks for the LMC and the SMC are briefly summarized in Table 1. Short term dynamical evolution ($\sim$0.3 Gyr) of the LMC and the SMC is investigated in each simulation. Description of the model parameters for the best model of the last LMC-SMC interaction are shown in table 3.

\subsection{The initial conditions}
In the best model presented in the present paper, the present-day distance of the LMC and the SMC is assumed to be 23 kpc as adopted in our previous simulations (Diaz \& Bekki 2012). The orbital plane of the SMC is inclined by 60 degrees with respect to the LMC disk and the SMC disk is included by 45 degrees with respect to the orbital plane. In these configurations of the orbits, the SMC gas disk can collide with the LMC gas disk with the pericenter distance of 4.2 kpc, which is well within the size of the LMC disk. It should be stressed here that the pericenter distance should be less than 5 kpc to cause gas accretion from the SMC to the LMC in the present study. As later described, this rapid accretion of the originally SMC gas is the essential ingredient of the L-component formation in the LMC disk. Such gas accretion from the SMC to the LMC was originally demonstrated in our previous simulations (Bekki \& Chiba 2007b), however, the exact locations of the gas accretion were not clearly described in these previous works. We therefore focus exclusively on the multiple locations of such gas accretion events within the LMC gas disk in the present study.

\begin{figure*}[htbp]
\begin{center}
\includegraphics[width=16cm]{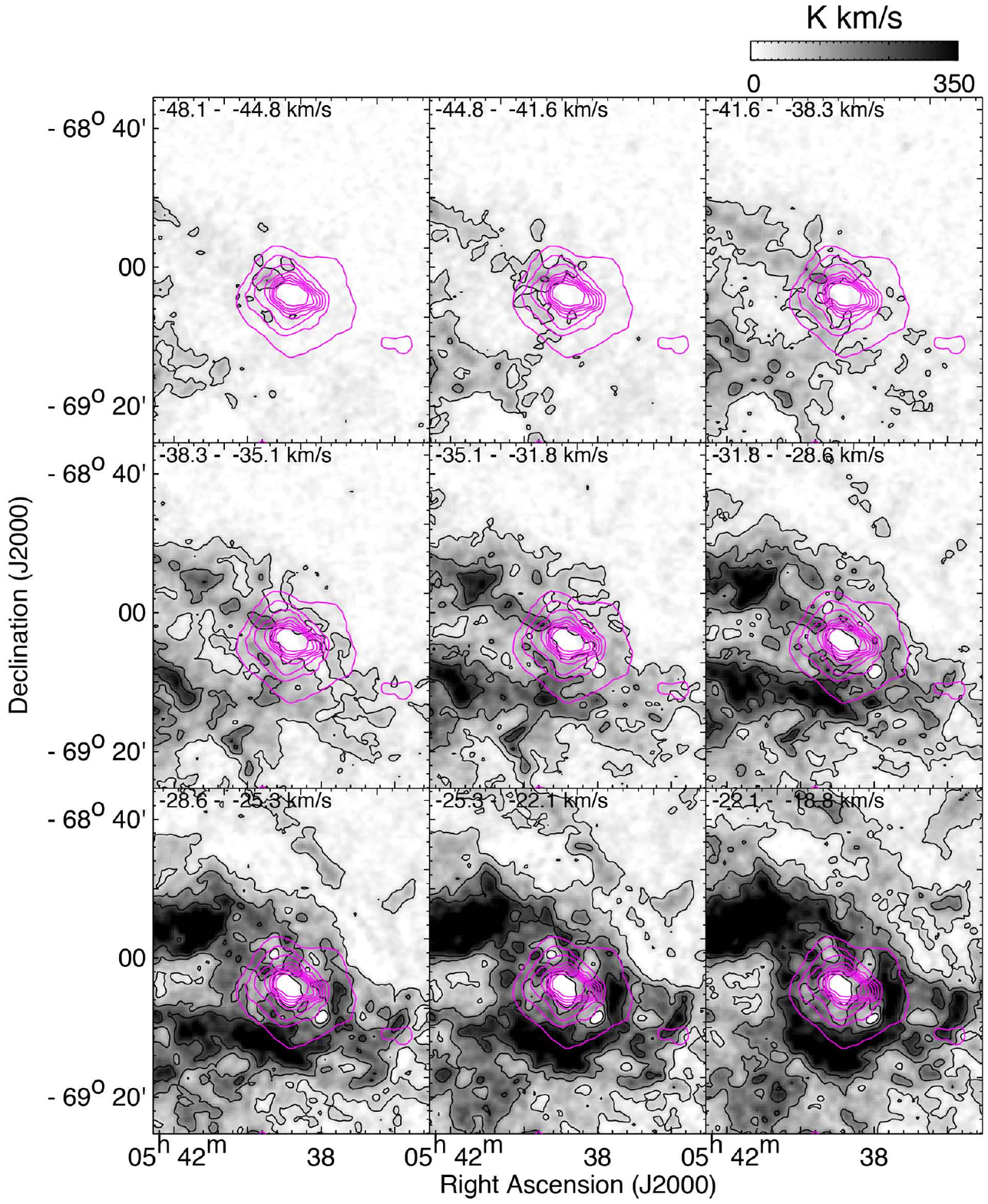}
\end{center}
\caption{Velocity channel maps of H{\sc i} gas with a velocity step of 3.25 km s$^{-1}$ overlaid with H$\alpha$ emission by magenta contours. The lowest level and intervals are 50 K km s$^{-1}$ and 100 K km s$^{-1}$ for H{\sc i} and 150 Rayleigh and 150 Rayleigh for H$\alpha$}  
\label{fig17}
\end{figure*}%

\setcounter{figure}{16}
\begin{figure*}[htbp]
\begin{center}
\includegraphics[width=16cm]{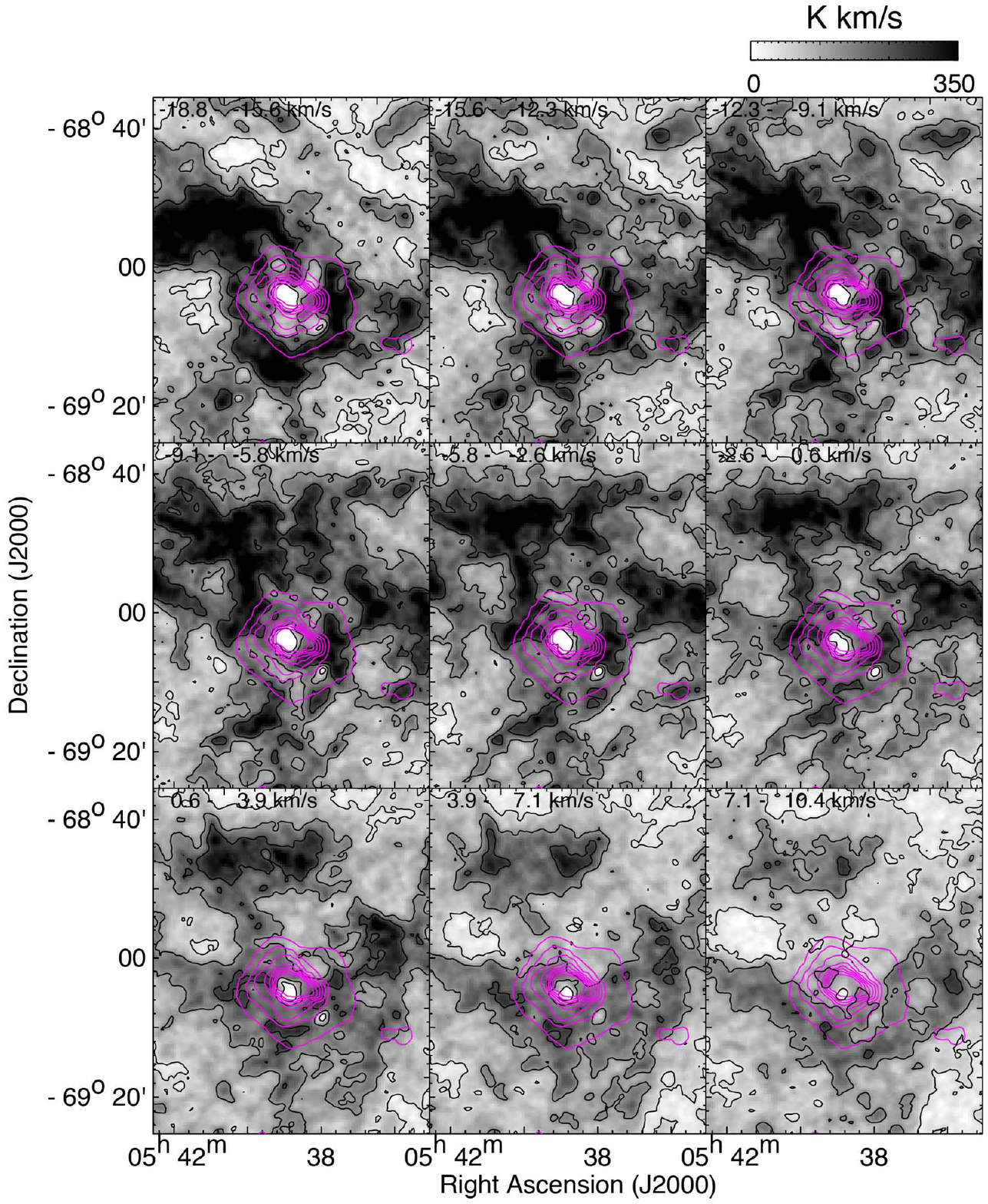}
\end{center}
\caption{Continued.}  
\label{fig17}
\end{figure*}%

\setcounter{figure}{16}
\begin{figure*}[htbp]
\begin{center}
\includegraphics[width=16cm]{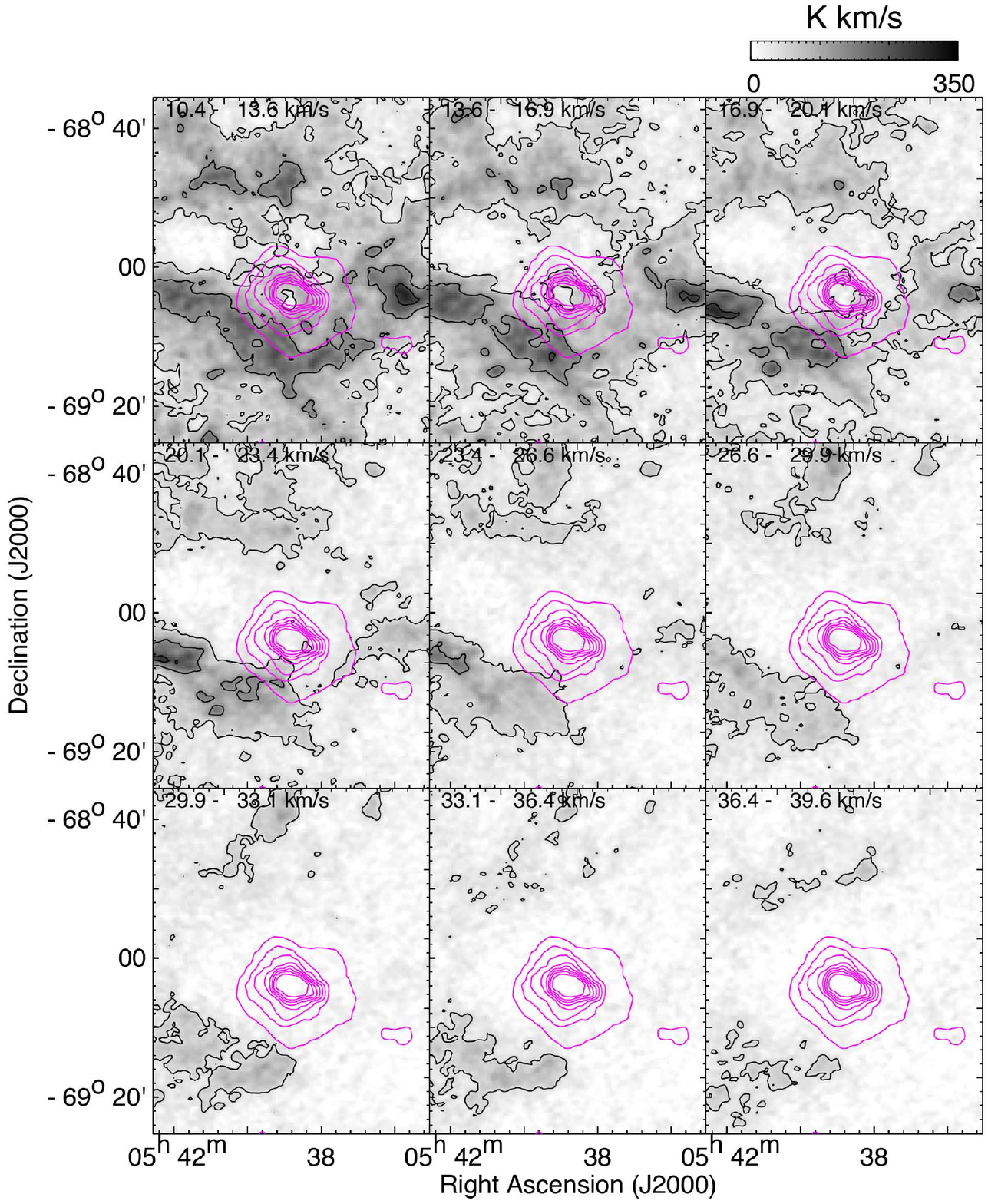}
\end{center}
\caption{Continued}  
\label{fig17}
\end{figure*}%

\begin{figure*}[htbp]
\begin{center}
\includegraphics[width=16cm]{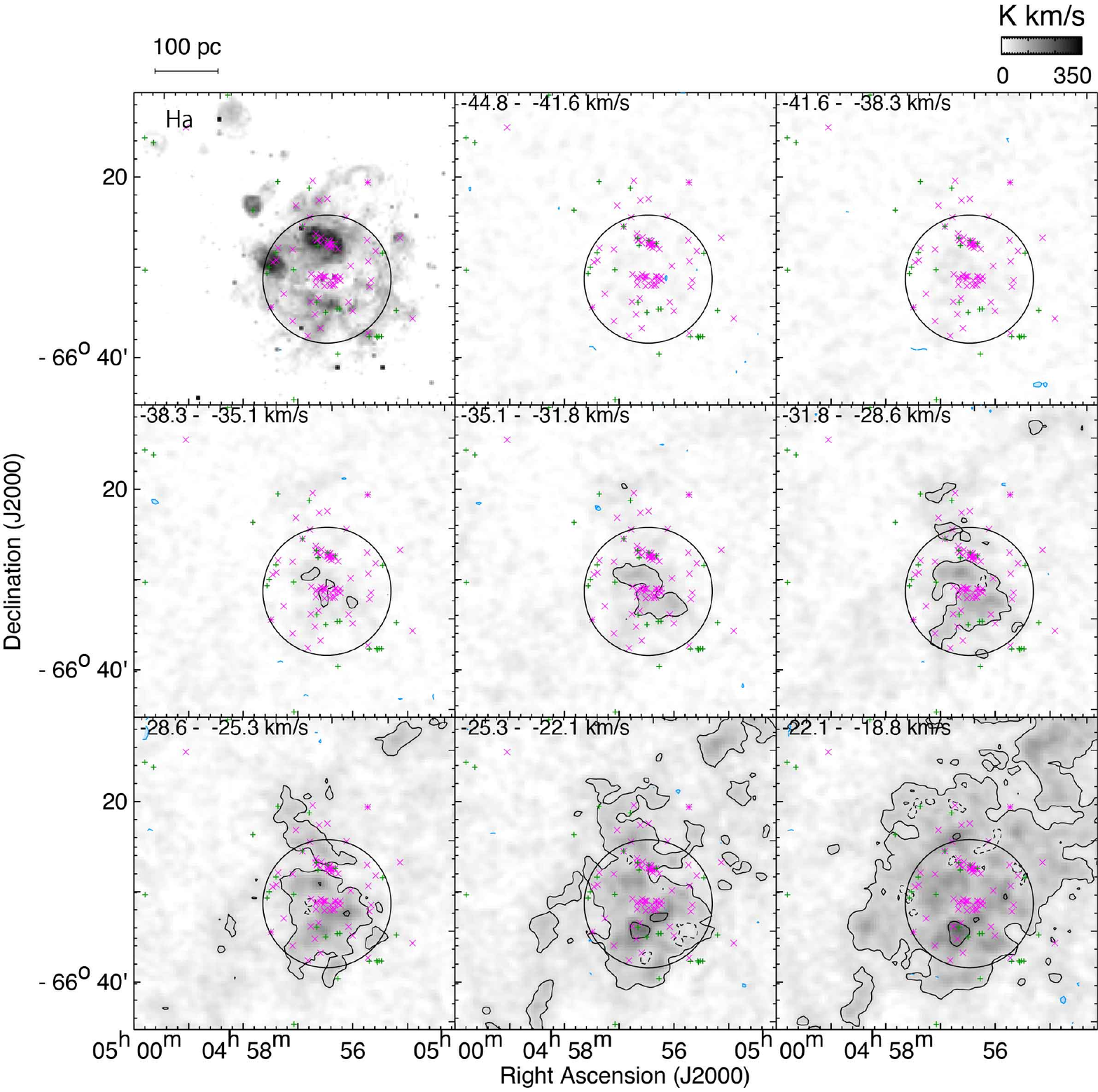}
\end{center}
\caption{Velocity channel maps of H{\sc i} gas with a velocity step of 3.25 km s$^{-1}$. The lowest level and intervals are 50 K km s$^{-1}$ and 100 K km s$^{-1}$.  The black circle indicates a ring morphology with a cavity of $\sim$100 pc in radius, enclosing OB association LH9 (Lucke \& Hodge 1970). Magenta asterisks and crosses are WR-stars and O-type stars, respectively (Bonanos et al. 2009). The green crosses indicate young stellar objects cataloged by Seale et al. (2009).}  
\label{fig18}
\end{figure*}%

\setcounter{figure}{17}
\begin{figure*}[htbp]
\begin{center}
\includegraphics[width=16cm]{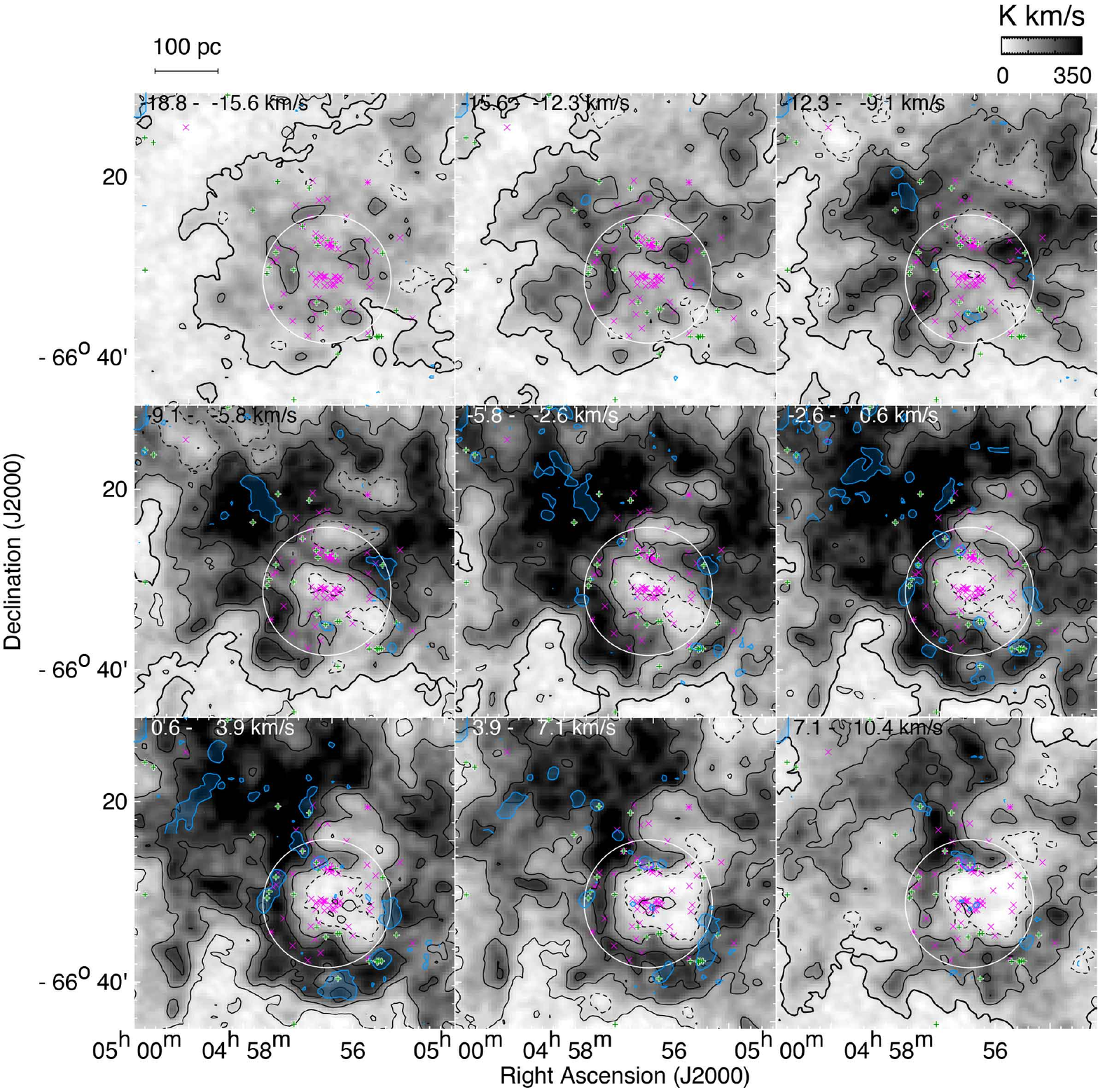}
\end{center}
\caption{Continued. The blue contours indicate velocity channel maps of $^{12}$CO ($J$=1--0) obtained with Mopra telescope (Wong et al. 2011). The contour levels is $\sim$1.2 K km s$^{-1}$ (3 $\sigma$). }  
\label{fig18}
\end{figure*}%


\setcounter{figure}{17}
\begin{figure*}[htbp]
\begin{center}
\includegraphics[width=16cm]{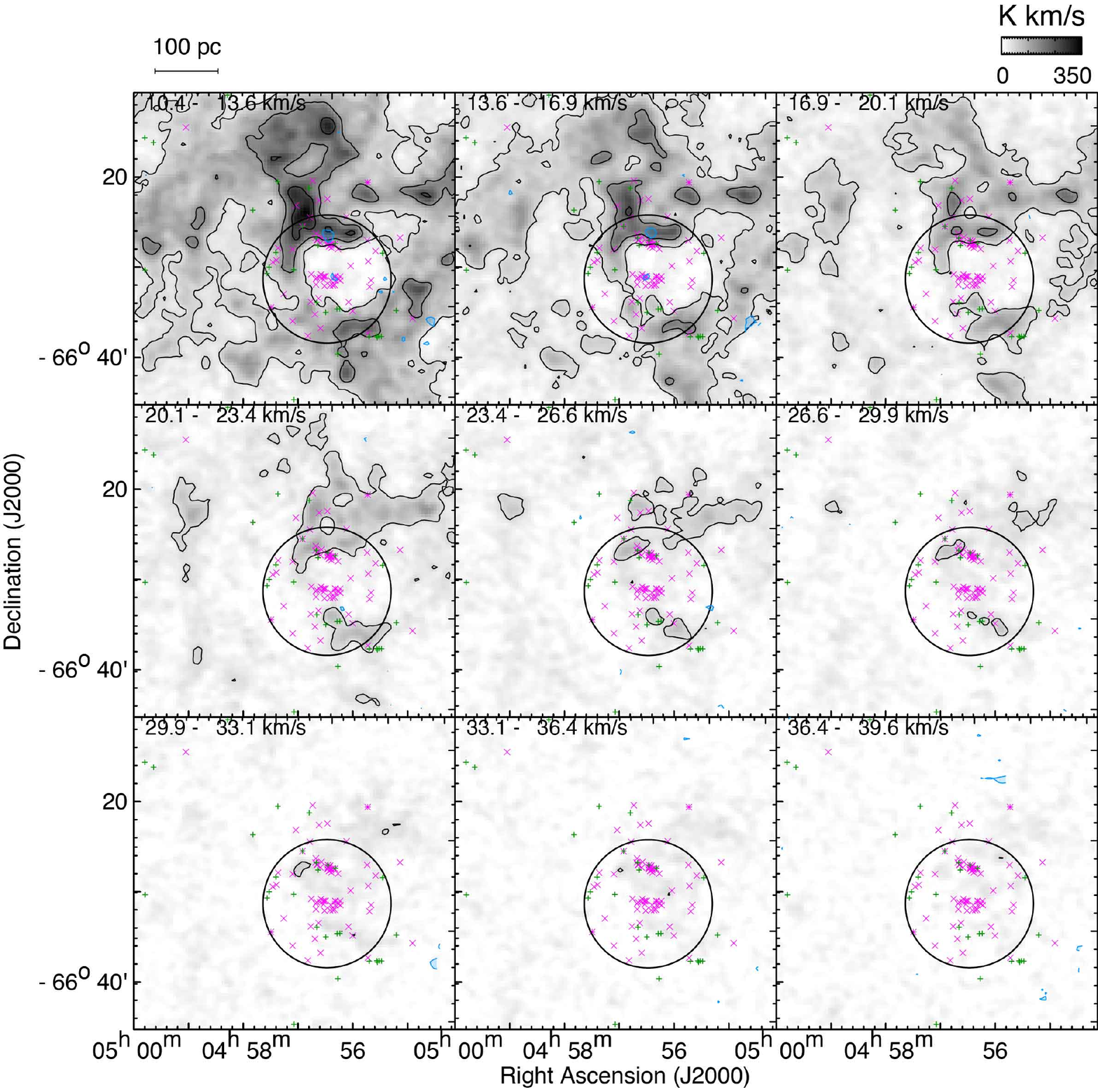}
\end{center}
\caption{Continued}  
\label{fig18}
\end{figure*}%
\clearpage

\begin{figure*}[htbp]
\begin{center}
\includegraphics[width=16cm]{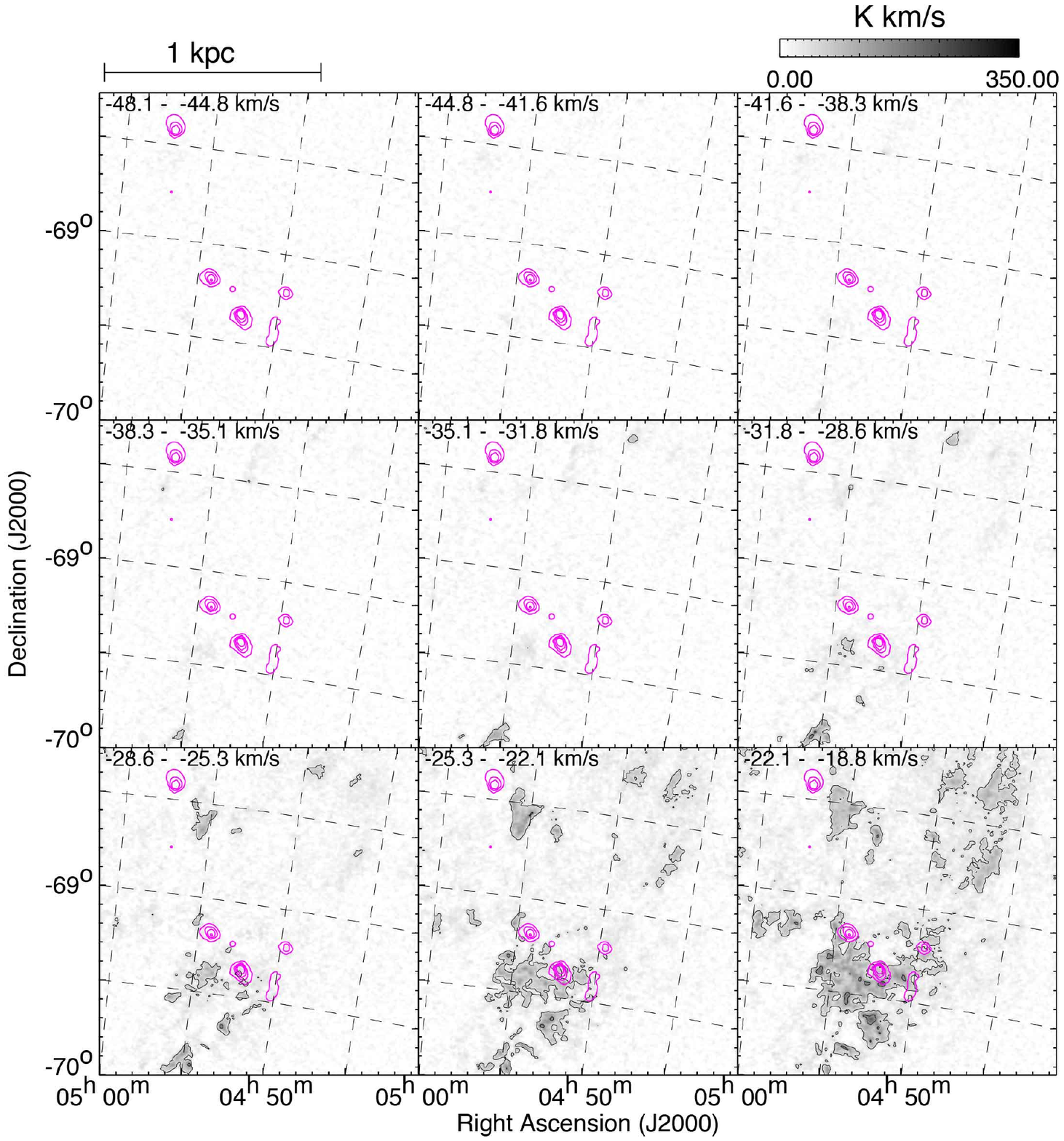}
\end{center}
\caption{Velocity channel maps of H{\sc i} gas with a velocity step of 3.25 km s$^{-1}$ overlaid with H$\alpha$ emission by magenta contours. The lowest level and intervals are 50 K km s$^{-1}$ and 100 K km s$^{-1}$ for H{\sc i} and 150 Rayleigh and 150 Rayleigh for H$\alpha$}  
\label{fig19}
\end{figure*}%

\setcounter{figure}{18}
\begin{figure*}[htbp]
\begin{center}
\includegraphics[width=16cm]{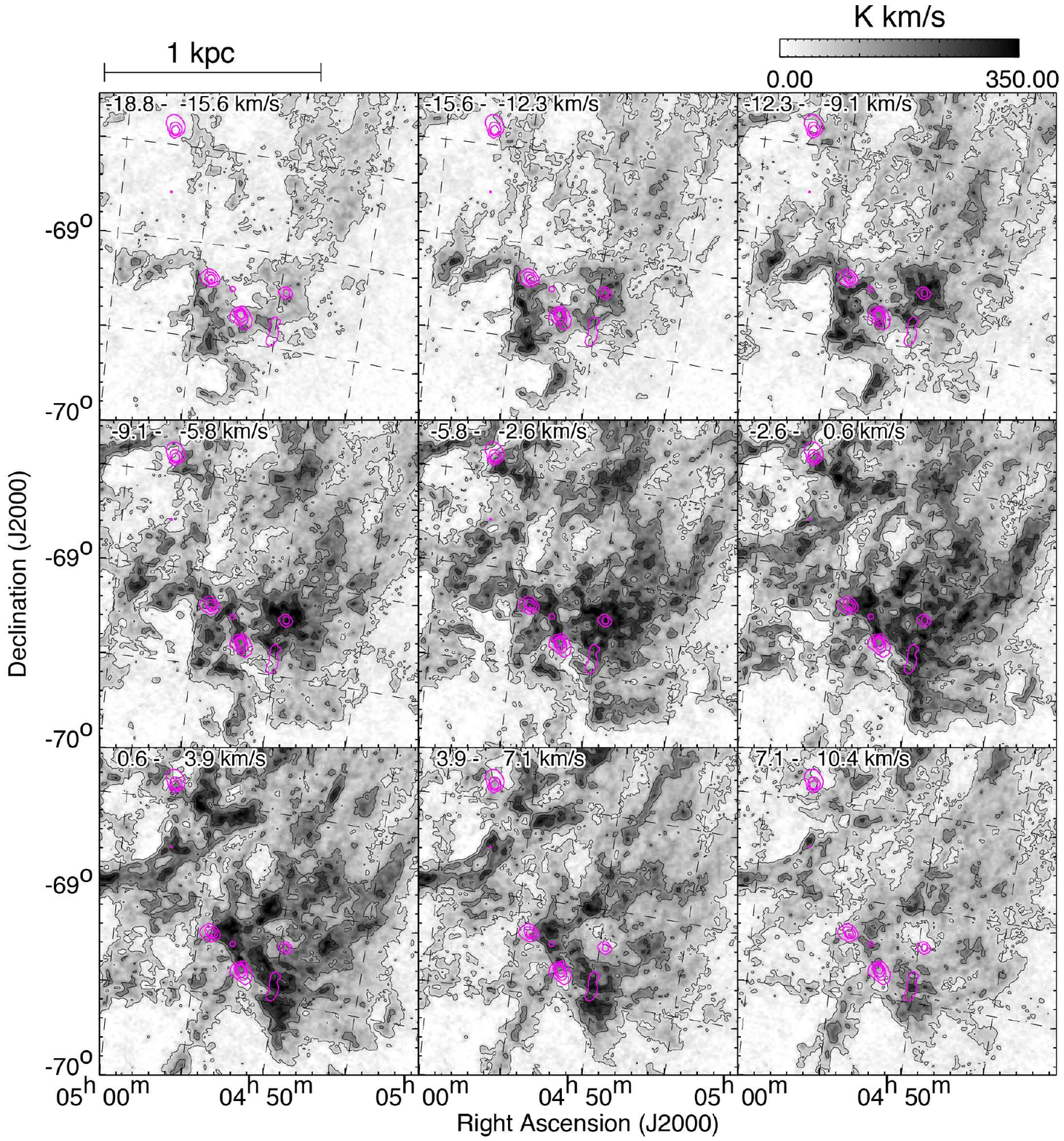}
\end{center}
\caption{Continued}  
\label{fig19}
\end{figure*}%

\setcounter{figure}{18}
\begin{figure*}[htbp]
\begin{center}
\includegraphics[width=16cm]{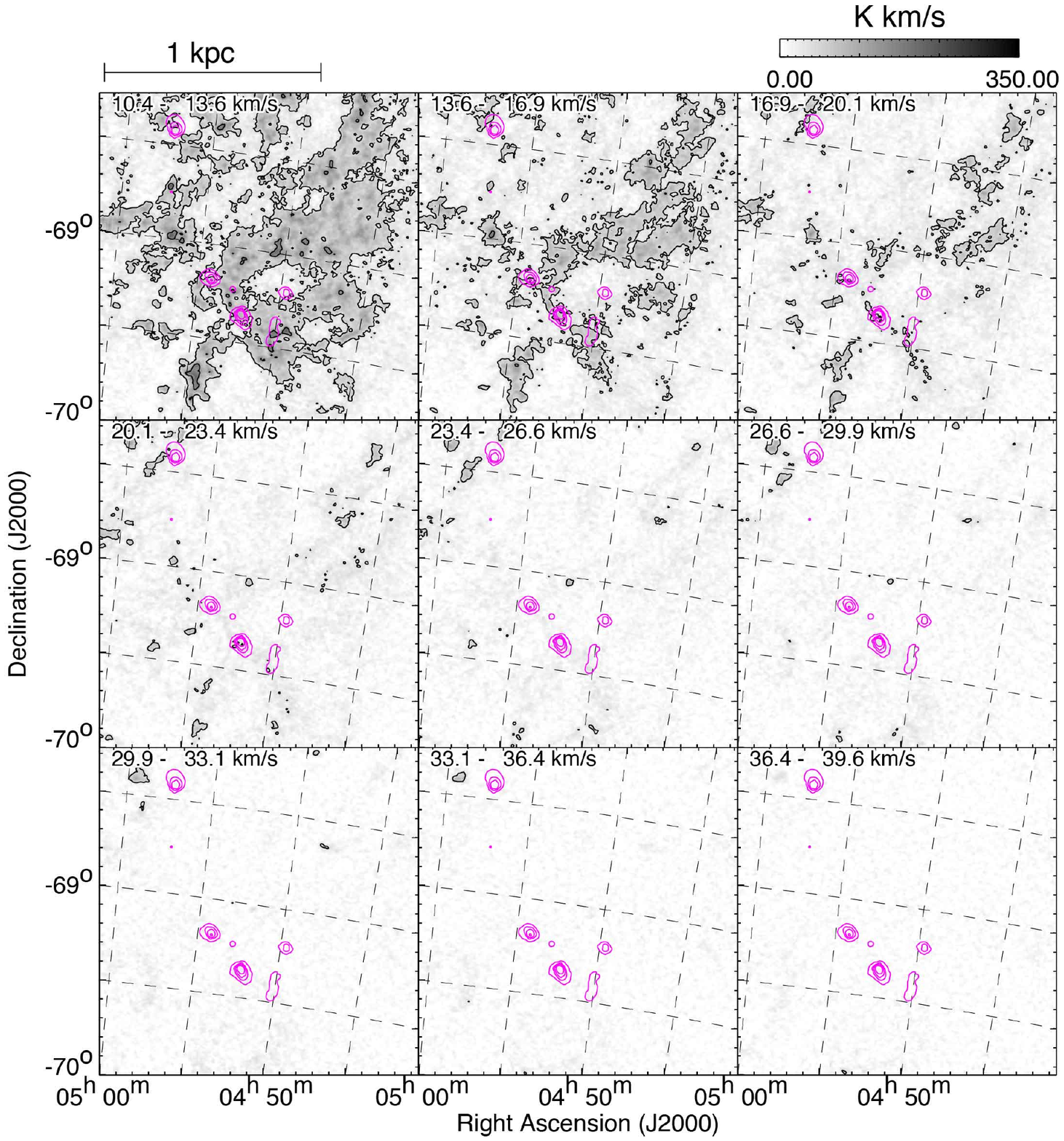}
\end{center}
\caption{Continued }  
\label{fig19}
\end{figure*}%

\setcounter{figure}{19}
\begin{figure*}[htbp]
\begin{center}
\includegraphics[width=\linewidth]{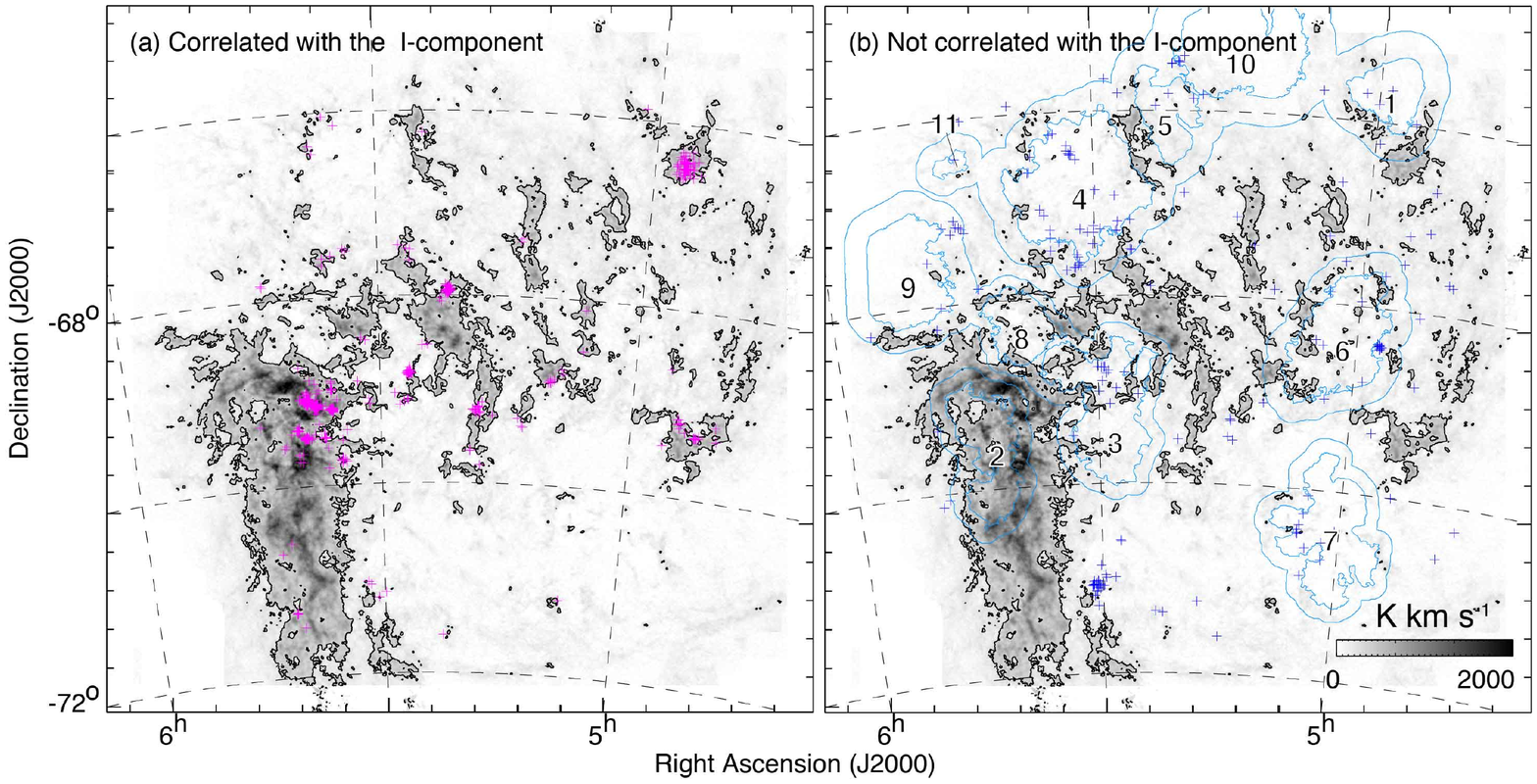}
\end{center}
\vspace*{-3cm}
\caption{(a) The distribution of high-mass stars correlated with the I-component by magenta crosses, which is corresponding to the blue histogram where integrated intensity is larger than 300 K km s$^{-1}$ in Figure 6(b). (b) The distribution of high-mass stars not correlated with the I-component by blue crosses. Light blue contours indicate the positions of super giant shells (Dawson et al. 2003). }  
\label{fig20}
\end{figure*}%

\setcounter{figure}{20}
\begin{figure*}[htbp]
\begin{center}
\includegraphics[width=15cm]{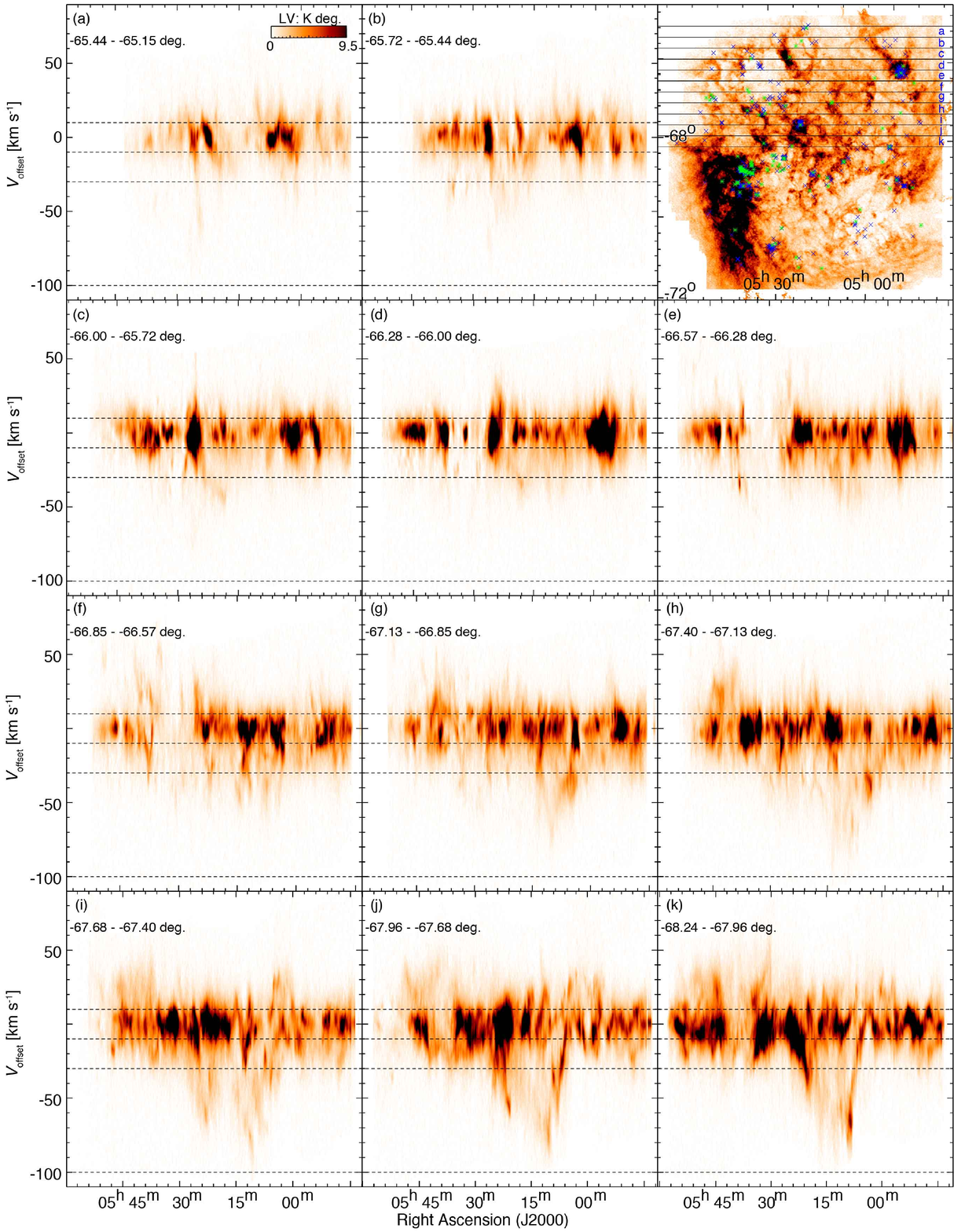}
\end{center}
\caption{The upper right panel shows total integrated intensity map of the LMC. {The green asterisks and blue crosses indicate WR stars and O-type stars (Bonanos et al. 2009)}. Horizontal lines indicate the integration ranges of Right Ascension--velocity diagrams in Dec. (a)--(k) Channel maps of Right Ascension--velocity diagrams over the whole H{\sc i}. The integration range is 0.27 deg. ($\sim$235.6 pc), and the integration range is shifted from north to south in 0.27 deg. step. The black dashed lines indicate the velocity ranges of the L-, I-, and D-components. The upper left number denotes the integration range in the upper right panel.}  
\label{fig21}
\end{figure*}%

\setcounter{figure}{20}
\begin{figure*}[htbp]
\begin{center}
\includegraphics[width=15cm]{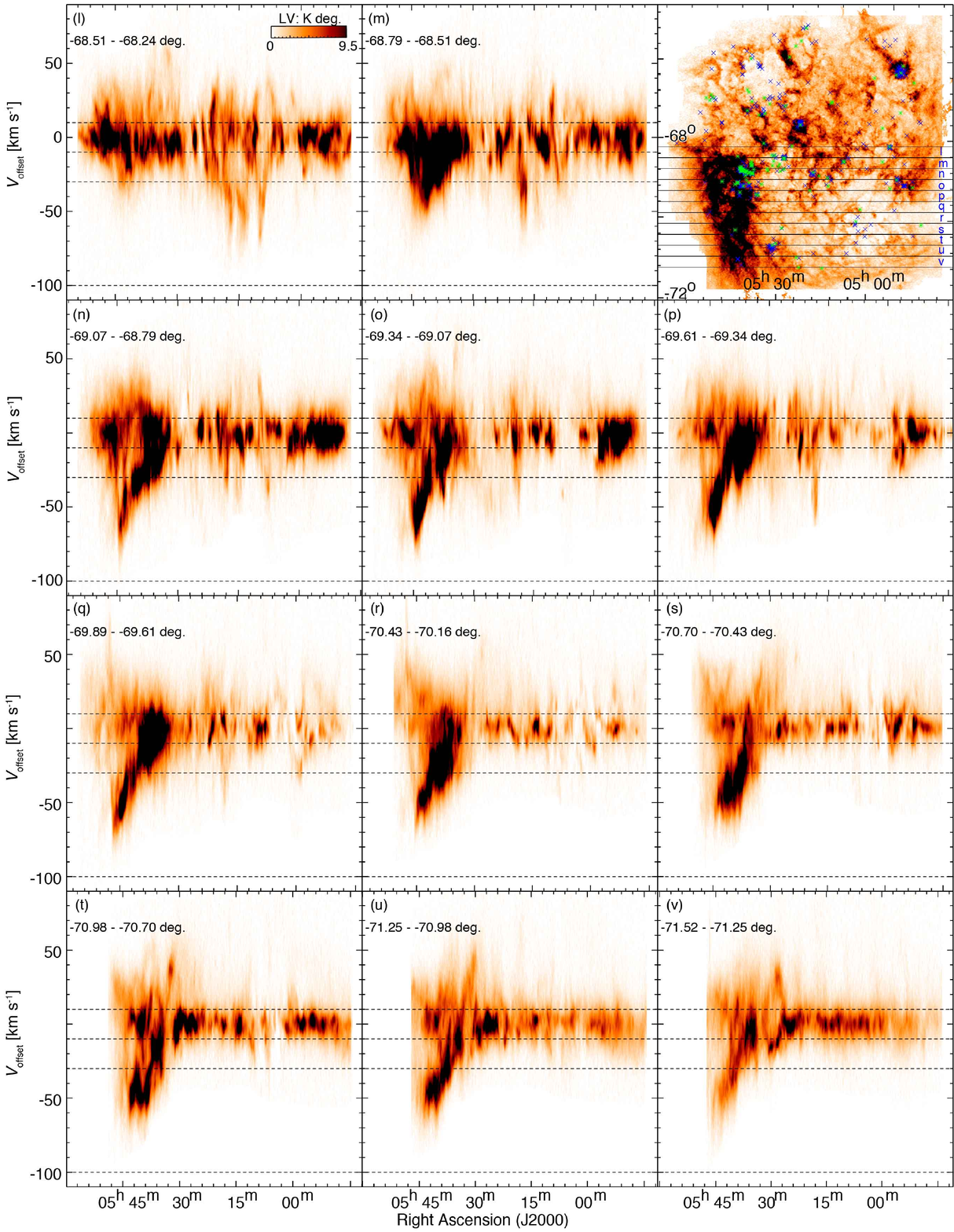}
\end{center}

\caption{Continued.}  
\label{fig21}
\end{figure*}%

\begin{deluxetable}{ccc}
\tablewidth{7.0cm}
\tablecaption{Description of the model parameters for the best model of the last LMC-SMC interaction.}
\label{tab:a1}
\tablehead{\multicolumn{1}{c}{Physical properties } &Parameter values }
\startdata
Total  halo mass (LMC)& 1.0$\times$10$^{11}$ $M_{\odot}$  \\
DM structure (LMC)& NFW profile \\ 
Virial radius (LMC) & $R_{\rm vir}$ = 77 kpc\\ 
$c$ parameter (LMC)&$c$=12\\
Stellar disk mass (LMC)& 3.4$\times$10$^9$ $M_{\odot}$\\
Gas disk mass (LMC)&1.4$\times$10$^9$ $M_{\odot}$\\
Stellar disk size (LMC)&5.5 kpc\\
Gas disk size (LMC)&5.5 kpc\\
Total halo mass (SMC)&1.0$\times$10$^{10}$ $M_{\odot}$\\
DM structure (SMC)&NFW profile\\
Virial radius (SMC)&$R_{\rm vir}$ = 24 kpc\\
$c$ parameter (SMC)&$c$=12\\
Stellar disk mass (SMC)&3.4$\times$10$^8$ $M_{\odot}$\\
Gas disk mass (SMC)&1.4$\times$10$^8$ $M_{\odot}$\\
Stellar disk size (SMC)&1.7 kpc\\
Gas disk size (SMC)& 1.7 kpc\\
Mass resolution&6.0$\times$10$^3$ $M_{\odot}$\\
size resolution&40 pc\\
LMC-SMC Mass-ratio&0.1\\
Pericenter distance of the two&4.2 kpc\\
Present-day distance of the two&23 kpc\\
\enddata
\tablecomments{}
\end{deluxetable}

\end{document}